\documentclass[fleqn,usenatbib,useAMS]{mnras}

\usepackage{caption}
\usepackage{subcaption}
\usepackage{graphicx}	
\usepackage{amsmath}	
\usepackage{amssymb}	
\usepackage{multicol}        
\usepackage{bm}		
\usepackage{pdflscape}	
\usepackage{afterpage}
\usepackage{xcolor,listings}
\usepackage{enumerate}
\usepackage{lastpage}

\DeclareMathOperator*{\maxi}{max}
\DeclareMathOperator*{\mini}{min}
\newcommand{\s}[1]{s_\mathrm{#1}}

\newcommand{\new}[1]{#1} 

\usepackage[T1]{fontenc}
\usepackage{ae,aecompl}

\usepackage{newtxtext,newtxmath}

\usepackage{graphicx}
\graphicspath{ {./images/} }
\usepackage{fontawesome5}
\usepackage{hyperref}
\usepackage{xurl}

\definecolor{orcidlogocol}{HTML}{A6CE39}

\hypersetup{colorlinks=true,
            citecolor=[rgb]{0, .4, .8},
            linkcolor=[rgb]{0, .4, .8},
            urlcolor=[rgb]{0, .4, .8}}
\title[Detecting Anomalous Galaxies with GANs]{Anomaly detection in Hyper Suprime-Cam galaxy images with generative adversarial networks}

\author[Storey-Fisher et al.]{Kate Storey-Fisher\ \href{http://orcid.org/0000-0001-8764-7103}{\textcolor{orcidlogocol}{\faOrcid}}$^{1}$\thanks{Contact e-mail: \href{mailto:k.sf@nyu.edu}{k.sf@nyu.edu}},
Marc Huertas-Company\ \href{http://orcid.org/0000-0002-1416-8483}{\textcolor{orcidlogocol}{\faOrcid}}$^{2,3}$,
Nesar Ramachandra\ \href{http://orcid.org/0000-0001-7772-0346}{\textcolor{orcidlogocol}{\faOrcid}}$^{4}$,
\newauthor
Francois Lanusse\ \href{http://orcid.org/0000-0001-7956-0542}{\textcolor{orcidlogocol}{\faOrcid}}$^{5}$,
Alexie Leauthaud\ \href{http://orcid.org/0000-0002-3677-3617}{\textcolor{orcidlogocol}{\faOrcid}}$^{6}$, 
Yifei Luo\ \href{http://orcid.org/0000-0001-7729-6629}{\textcolor{orcidlogocol}{\faOrcid}}$^{6}$,
Song Huang\ \href{http://orcid.org/0000-0003-1385-7591}{\textcolor{orcidlogocol}{\faOrcid}}$^{7}$,
J. Xavier Prochaska\ \href{http://orcid.org/0000-0002-7738-6875}{\textcolor{orcidlogocol}{\faOrcid}}$^{6,8}$
\\
$^{1}$ Center for Cosmology and Particle Physics, Department of Physics, New York University, NY 10003, USA\\
$^{2}$Instituto de Astrof\'isica de Canarias (IAC); Departamento de Astrof\'isica, Universidad de La Laguna (ULL), E-38200, La Laguna, Spain\\
$^{3}$LERMA, PSL Research University, Observatoire de Paris, CNRS, Sorbonne Universites, UPMC Univ. Paris 06, F-75014 Paris, France\\
$^{4}$High Energy Physics Division, Argonne National Laboratory, Lemont, IL 60439, USA\\
$^{5}$CPS Division, Argonne National Laboratory, 9700 South Cass Avenue, Lemont, IL 60439, USA\\
$^{6}$AIM, CEA, CNRS, Universit\'e Paris-Saclay, Universit\'e de Paris, F-91191 Gif-sur-Yvette, France\\
$^{7}$Department of Astronomy and Astrophysics, University of California, Santa Cruz, Santa Cruz, CA 95064, USA \\
$^{8}$Department of Astrophysical Sciences, Princeton University, Princeton, NJ 08544, USA\\
$^{9}$Kavli Institute for the Physics and Mathematics of the Universe (Kavli IPMU), 5-1-5 Kashiwanoha, Kashiwa, 277-8583, Japan\\
}
\date{Last updated 2021 September}

\pubyear{2021}

\begin{document}
\label{firstpage}
\pagerange{\pageref{firstpage}--\pageref{lastpage}}
\maketitle

\begin{abstract}
The problem of anomaly detection in astronomical surveys is becoming increasingly important as data sets grow in size.
We present the results of an unsupervised anomaly detection method using a Wasserstein generative adversarial network (WGAN) on nearly one million optical galaxy images in the Hyper Suprime-Cam (HSC) survey.
The WGAN learns to generate realistic HSC-like galaxies that follow the distribution of the data set; anomalous images are defined based on a poor reconstruction by the generator and outlying features learned by the discriminator.
We find that the discriminator is more attuned to potentially interesting anomalies compared to the generator, \new{and compared to a simpler autoencoder-based anomaly detection approach}, so we use the discriminator-selected images to construct a high-anomaly sample of $\sim$13,000 objects.
We propose a new approach to further characterize these anomalous images: we use a convolutional autoencoder to reduce the dimensionality of the residual differences between the real and WGAN-reconstructed images and perform UMAP clustering on these.
We report detected anomalies of interest including galaxy mergers, tidal features, and extreme star-forming galaxies.
\new{A follow-up spectroscopic analysis of one of these anomalies is detailed in the Appendix; we find that it is an unusual system most likely to be a metal-poor dwarf galaxy with an extremely blue, higher-metallicity HII region.}
We have released a catalog with the WGAN anomaly scores; the code and catalog are available at \url{https://github.com/kstoreyf/anomalies-GAN-HSC}, and our interactive visualization tool for exploring the clustered data is at \url{https://weirdgalaxi.es}.

\end{abstract}




\section{Introduction}

Many discoveries in astronomy have been made by identifying unexpected outliers in collected data (e.g. \citealt{Cardamone2009}, \citealt{Massey2019}). 
These outliers, also referred to as anomalies or novelties, are data points that lie outside of the normal distribution of data.
In the astronomy context, we are interested in finding unknown classes of objects, objects belonging to rare classes, and individual objects of known type with anomalous properties.
As data sets increase in size, automated methods for detecting these outliers are becoming necessary.
The Sloan Digital Sky Survey (SDSS) surveyed one third of the sky and observed over 1 billion cataloged objects \citep{York2000}.
In the near future, the Rubin Observatory will observe 40 billion objects \citep{Ivezic2018}.
These present opportunities for novel discoveries in their massive data sets, as well as the need for new, automated methods to filter the data and identify anomalies.

Outlier identification has been an area of study since as early as the 19th century \citep{Edgeworth1887}.
\new{This early work was primarily focused on identifying and filtering out measurement errors and other non-interesting outliers.
While this is still a main avenue in modern work on anomaly detection, and is in fact increasingly important for today's large data sets, we have become interested using in anomaly detection for discovery.}
Recent work in astronomy along these lines has applied a range of statistical and computational techniques.
A nearest neighbors approach, often combined with a dimensionality reduction step, has been used for outlier detection in cross-matched astronomical data sets \citep{Henrion2013}.
Applications often target specific types of objects, such as using Bayesian model selection to select rare high-redshift quasars from a star-dominated population \citep{Mortlock2012}.
Another approach is Principal Component Analysis (PCA) to identify distinguishing features; for instance, \cite{Dutta2007} used PCA for anomaly detection in SDSS and 2MASS flux and surface brightness data.

Machine learning methods are being rapidly developed as approaches to anomaly detection in astronomy and other fields.
A review of anomaly detection methods and applications using deep learning is presented in \cite{Chalapathy2019}.
Unsupervised learning lends itself to this problem, as it allows for outlier identification without expert labelling of training data or introducing biases based on expected outliers.
\cite{Baron2017} use random forests to find outliers in Sloan Digital Sky Survey (SDSS) spectroscopic data.
\cite{Solarz2017} apply support vector machines to find anomalies in the Wide-field Infrared Survey Explorer (WISE) survey.
\cite{Segal2019} explore the use of apparent complexity as a feature for machine learning algorithms to better identify radio galaxies with complex morphology.
Unsupervised learning has also been applied to anomaly detection problems beyond galaxy surveys, including on supernovae data \citep{Pruzhinskaya2019} and Kepler light curves \citep{Giles2019}.
Finally, general frameworks for anomaly detection have been developed, such as the combined machine learning--human input approach of \cite{Lochner2020}.

Deep generative models present another class of approaches to anomaly detection.
These have a natural application to identifying outliers, as they are able to model complex distributions of high-dimensional data.
\new{Autoencoders have recently shown promise in this realm:
\cite{Morawski2021} apply a convolutional autoencoder (CAE) to gravitational wave data, and \cite{DAddona2021} use a disentangled CAE to find outliers in KiDS data.
Variational autoencoders (VAEs) are a natural choice for anomaly detection as they provide a direct probability measure \citep{An2015}; \cite{Villar2021} use a VAE to detect anomalous extragalactic transients.
However, vanilla CAEs are simple models that do not always reproduce the data well, and VAEs are known to suffer from constraining priors and over-regularization \citep{Ghosh2020}, limiting their application to anomaly detection.}

Another deep generative model class that is gaining popularity is generative adversarial networks (GANs), proposed by \cite{Goodfellow2014}.
GANs were first applied to anomaly detection by \cite{Schlegl2017}, in the context of medical imaging.
They demonstrate that a GAN trained on normal images can then be used to identify abnormal images.
\cite{Zenati2018} show that training an encoder simultaneously with the GAN improves testing efficiency, and they demonstrate their performance on outlier detection tasks on a range of high-dimensional data.
GANs have also been used to detect outliers in time-series data \citep{Li2018}.
Recently, \cite{Margalef-Bentabol2020} used a GAN to detect merging galaxies and compare galaxy simulations against observations. 
\cite{DiMattia2019} present a survey of the application of GANs to anomaly detection and perform empirical validation of the models.

The Hyper Suprime-Cam Subaru Strategic Program (HSC-SSP) is a natural data set for anomaly detection applications \citep{Miyazaki2018}.
It is a wide-field optical survey with very good seeing, an average of 0.6 (FWHM) in the $i$-band, and a deep magnitude limit, of 26.2 given a $5\sigma$ point source detection limit.
Many interesting objects have already been identified in HSC; some have been found by machine learning algorithms, such as in the case of galaxy interaction signatures \citep{Goulding2017}, while others used more traditional selection techniques targeted to the objects of interest, such as for strong gravitational lenses \citep{Wong2018} and extended emission line objects \citep{Sun2018}.
We are interested in finding more of these types of objects, as well as galaxies with extreme colors, galaxies with extreme activity, rare quasars, and other rare or interesting objects.
In addition to these, anomaly detection will be useful for filtering out optical artifacts in HSC.

In this work, we train a GAN to identify anomalous objects in a subsample of galaxies with HSC imaging.
We then characterize the anomalous images with a new convolutional autoencoder-based approach, and identify a set of scientifically interesting anomalies.
\new{We also compare the GAN-detected anomalies to those found with a simple CAE approach.}
This paper is organized as follows.
In Section~\ref{sec:data}, we detail the galaxy image data set used in our application.
We describe our WGAN model, approach to anomaly score assignment, technique for characterization, \new{and CAE model} in Section~\ref{sec:model}.
In Section~\ref{sec:results} we discuss our results and show anomalous galaxies identified with our framework.
We present a summary and our conclusions in Section~\ref{sec:conclusions}.
\new{A follow-up spectroscopic analysis of one of our GAN-detected anomalies is described in Appendix~\ref{sec:bluedot}.}

\section{Data}
\label{sec:data}

\subsection{Hyper Suprime-Cam Survey}

We use data from the Hyper Suprime-Cam Subaru Strategic Program \citep{Aihara2018a}.
The wide-field optical survey is imaged with the Subaru Telescope and has been ongoing since March 2014, with the first public data release in 2018 \citep{Aihara2018b}.
HSC provides extremely high sensitivity and resolving power, thanks to the large 8.2 meter mirror of the Subaru Telescope; its $i$-band seeing of 0.6 (FWHM) is a large improvement from SDSS, which has a typical $i$-band seeing of 1.4.

We work with the second public data release (PDR2, \citealt{Aihara2019}), which contains over 430 million primary objects in the wide field covering 1114 deg$^2$. 
Of this wide field, a 305 deg$^2$ area is observed in full-depth full-color.
The objects in this field are observed in 5 broad-band filters, \textit{grizy}, with $5\sigma$ point-source detection limits within a 2 arcsec diameter aperture of 26.2 mag in the $r$- and $i$-bands, 26.6 mag in the $g$-band, and 25.4 mag in the $z$-band.

\subsection{Selection of Sample} 

The HSC data are reduced using a sophisticated data reduction pipeline \citep{Bosch2019}.
We use the force-photometry catalog for sample selection. 
Using the $i$-band CModel photometry, we select objects within a magnitude range of $20.0 < i < 20.5$ for our analysis. 
This choice of a thin magnitude slice allows for a more consistent sample in object size; we note that this choice is important for ease of analysis, as detection on a wide range of magnitudes runs the risk of a bias towards identifying bright objects as anomalies.
For application to larger samples, one could train separate WGANs on each slice, or carefully choose training batches to balance magnitudes; we leave this for future work.

We exclude objects flagged as having significant issues by the pipeline. 
These are, in any band: cosmic rays crossing the center pixel, saturated center pixel, interpolated center pixel, source at edge of survey volume, failed flux fit.
The full query, including these cuts and the information we retain about each sample, can be found at \url{https://github.com/kstoreyf/anomalies-GAN-HSC/blob/master/prepdata/hsc_pdr2_query.sql}, and can be run through the HSC data access site at \url{https://hsc-release.mtk.nao.ac.jp/datasearch}.

\begin{figure*}
    \begin{subfigure}{.4\textwidth}
    \centering
    \includegraphics[width=1\linewidth]{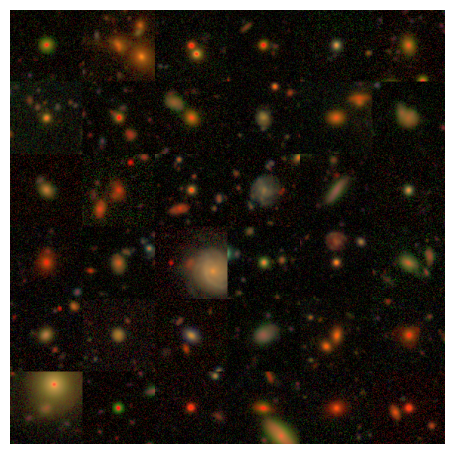}
    \caption{}
    \label{fig:real}
    \end{subfigure}
    \hspace{4em}
    \begin{subfigure}{.4\textwidth}
    \centering
    \includegraphics[width=1\linewidth]{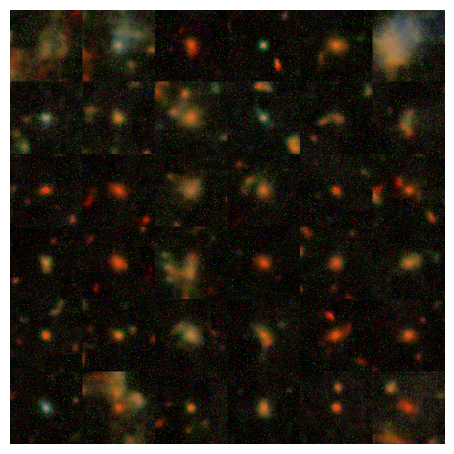}
    \caption{}
    \label{fig:gen}
    \end{subfigure}
\caption{(a) A random subsample of the HSC images used for training the WGAN and identifying anomalies. (b) A random sample of images generated by the WGAN, each conditioned on a latent-space vector drawn from a random normal distribution.}
\end{figure*}

We generate cutouts of $96 \times 96$ pixels ($\sim15\times 15$ arcsec) around each object; this captures the entirety of most objects while still being a reasonable size for training the network.
We use the $gri$-bands to construct 3-color images.
This results in a sample of 942,782 objects.
A random subsample from our final training data selection is shown in Figure \ref{fig:real}.
We see the sample contains many compact objects, most of which are faraway galaxies.
The red compact sources may be high-redshift galaxies, but some of them may also be stars; low-temperature dwarfs look similar in the survey, and are not simple to filter out.
We also see some more extended galaxies, including some with clear structure including bulges and spiral arms. 

We first preprocess the images to avoid issues due to the raw data range spanning multiple orders of magnitude.
We convert the flux values to RGB values using the method of \citealt{Lupton2004}, which rescales the data using the inverse hyperbolic sine function (asinh).
We use a stretch value of 0.5 and a softening parameter of 15.
This produces values from 0 to 255 for each pixel in each band, and we then normalize these image by image to between 0 and 1.
This scaling may affect the features identified by the WGAN as anomalous; for instance, it may suppress the degree of anomaly of galaxies that have extremely high or low flux in certain regions.
However, the rescaled images should largely retain the information about each object, which is sufficient for this work; we leave further exploration of the effect of flux conversions to future work.

\section{Model and Training}
\label{sec:model}

\subsection{WGAN Architecture and Training}

The standard GAN framework, introduced in \cite{Goodfellow2014}, involves a generator and a discriminator which compete against each other in a minimax game.
The discriminator $D(\bm{x})$ learns to distinguish real images $\bm{x}$ from those generated by the generator $G(\bm{z})$, where $\bm{z}$ is a vector in the generator's latent space.
The generator, in turn, learns how realistic its generated images are based on feedback from the discriminator.
The loss function to optimize is then
\begin{multline}
\mini_G \, \maxi_D \, L_{D,G} = \mathbb{E}_{\bm{x}\sim p_{\mathrm{r}}(\bm{x})}[\mathrm{log} (D(\bm{x}))] \: + \\ \mathbb{E}_{\bm{z}\sim p_{\mathrm{z}}(\bm{z})}[\mathrm{log}(1 - D(G(\bm{z})))] 
\end{multline}
where $\mathbb{E}_{\bm{x}\sim p_{\mathrm{r}}(\bm{x})}$ is the expectation value over the distribution of all real images, and $\mathbb{E}_{\bm{z}\sim p_{\mathrm{z}}(\bm{z})}$ is the expectation value over random input vectors to the generator.

GANs are notorious for instability in training; balancing the generator and discriminator losses is nontrivial, and vanilla GANs can fail to find a Nash equilibrium \citep{Salimans2016}.
One improvement is to use the Wasserstein distance as a metric to compare real and generated samples.
This is a more meaningful distance measure and is smooth even when the distributions are disjoint.
Wasserstein GANs are shown to be more stable and avoid vanilla GAN issues including mode collapse \citep{Arjovsky2017}.
The WGAN formulation is qualitatively different from that of vanilla GANs in that, rather than the discriminator outputting a probability of whether the sample was drawn from the real or generated distribution, the discriminator outputs the Wasserstein distance between the probability distributions of real and generated samples.
The WGAN discriminator is often called the ``critic'' to distinguish it, though we will continue to use ``discriminator'' for clarity.

The Wasserstein distance can be approximated by
\begin{equation}
    W\left(p_{\mathrm{r}}(\bm{x}\right), p_{\mathrm{z}}(\bm{z})) 
    = \text{sup}\left[ \mathbb{E}_{\bm{x}\sim p_{\mathrm{r}}(\bm{x})}[D(\bm{x})]
    - \mathbb{E}_{\bm{z}\sim p_{\mathrm{z}}(\bm{z})}[D(G(\bm{z}))]
    \right]
\end{equation}
where ``sup'' denotes the supremum over discriminator parameters. 
The generator aims to minimize this distance to achieve a distribution of generated images closest to the distribution of real images.
The generator only enters into the second term, so its loss function can be defined as
\begin{equation}
    L_G
    = - \mathbb{E}_{\bm{z}\sim p_{\mathrm{z}}(\bm{z})}[D(G(\bm{z}))] ~.
\end{equation}
The discriminator aims to maximize the Wasserstein distance; its loss function can be taken as the difference between the discriminator output on the real and generated samples.
However, a discriminator with this naive loss would get ``stuck'' when the discriminator is perfect, as the gradient vanishes and the loss function cannot continue to be updated.
One approach to address this is to apply a gradient penalty (GP) to penalize the loss to avoid vanishing gradients; to implement this we add a regularization term to the loss.
The discriminator loss is then
\begin{multline}
        L_D
    = \mathbb{E}_{\bm{x}\sim p_{\mathrm{r}}(\bm{x})}[D(\bm{x})]
    - \mathbb{E}_{\bm{z}\sim p_{\mathrm{z}}(\bm{z})}[D(G(\bm{z}))]
    \: + \\ \lambda_\text{GP} \, \mathbb{E}_{{\bm{y}}\sim p_{\mathrm{y}}} [(||\nabla_\bld{y}D(\bld{y})||_2 - 1)^2]
\end{multline}
where $\bm{y} = \epsilon \bm{x} + (1-\epsilon)G(\bm{z})$ is a uniform sampling from the line between samples from the real and generated distributions (with $\epsilon$ as a mixing parameter drawn uniformly from $0\leq\epsilon\leq1$), $\lambda_\text{GP}$ is the hyperparameter controlling the strength of the regularization, and $||\cdot||_2$ is the $L_2$ norm.

For our model we use this standard formulation of a Wasserstein GAN with gradient penalty (WGAN-GP).
The implementation is in \texttt{tensorflow} and \texttt{python}, based on that by \cite{Gulrajani2017}.
The generator and the discriminator are convolutional neural networks; the generator takes as input a latent space vector of dimension 128, and outputs an image of size $96\times96\times3$ pixels where 3 is the number of color bands.
It has a depth of 4 and a sigmoid activation function.
The discriminator takes as input an image of size $96\times96\times3$ pixels and outputs a real number; it also has a depth of 4.

We train the WGAN in batches of 32 images, with 5 discriminator updates per generator update.
For the loss functions we use the parameter $\lambda_\text{GP}=10$, and maximize the losses with the Adam optimizers with hyperparameters which we tune, choosing $\alpha=10^{-4}$, $\beta_1=0.5$, and $\beta_2=0.9$, for both the discriminator and generator.
We finalize the model at around 10,000 training iterations, after which the generator and discriminator losses stabilize and no longer improve.

A random sample of images generated with this WGAN, starting from latent space vectors drawn from a normal distribution, is shown in Figure \ref{fig:gen}.
We can see that the WGAN is able to generate realistic images for less extended objects, including neighboring objects and noise properties.
It also captures the color distribution of the population well, and reproduces the general noise properties.
However, it is unable to reconstruct the finer structure of more extended objects, such as spiral arms and distinct bulges; instead it generates diffuse-looking objects that do not always represent realistic galaxies.
That said, the fact that the WGAN does not learn to generate these more structured images is in line with our application of the WGAN to anomaly detection, as this limitation is precisely due to the fact that these images are less well-represented in the data.
We quantify the ability of the WGAN to represent the training objects and the connection to anomaly detection in the next section.

\subsection{Anomaly Score Assignment}
\label{sec:sanom_assignment}

\begin{figure*}
\begin{subfigure}{.325\textwidth}
  \centering
  \includegraphics[width=1\linewidth]{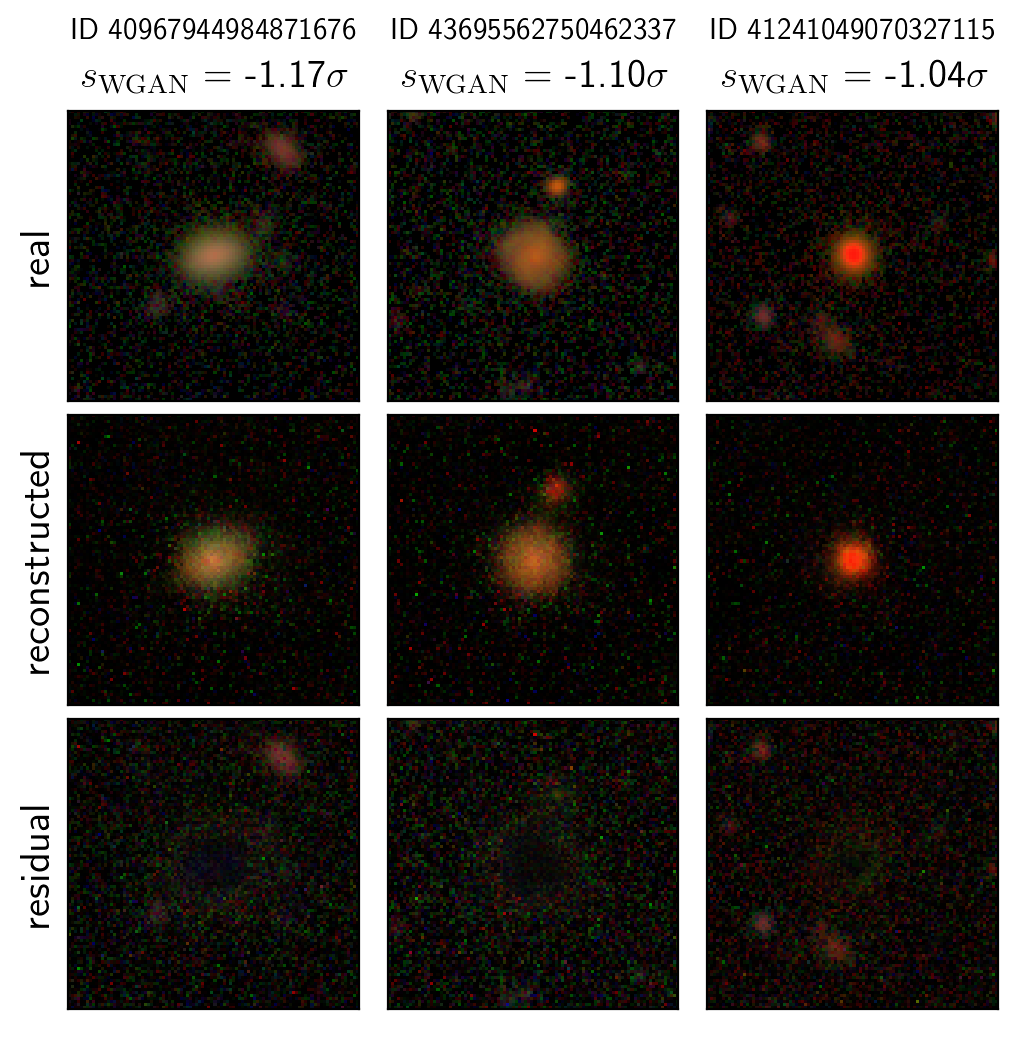}  
  \caption{}
  \label{fig:recon_neg}
\end{subfigure}
\hfill
\begin{subfigure}{.325\textwidth}
  \centering
  \includegraphics[width=1\linewidth]{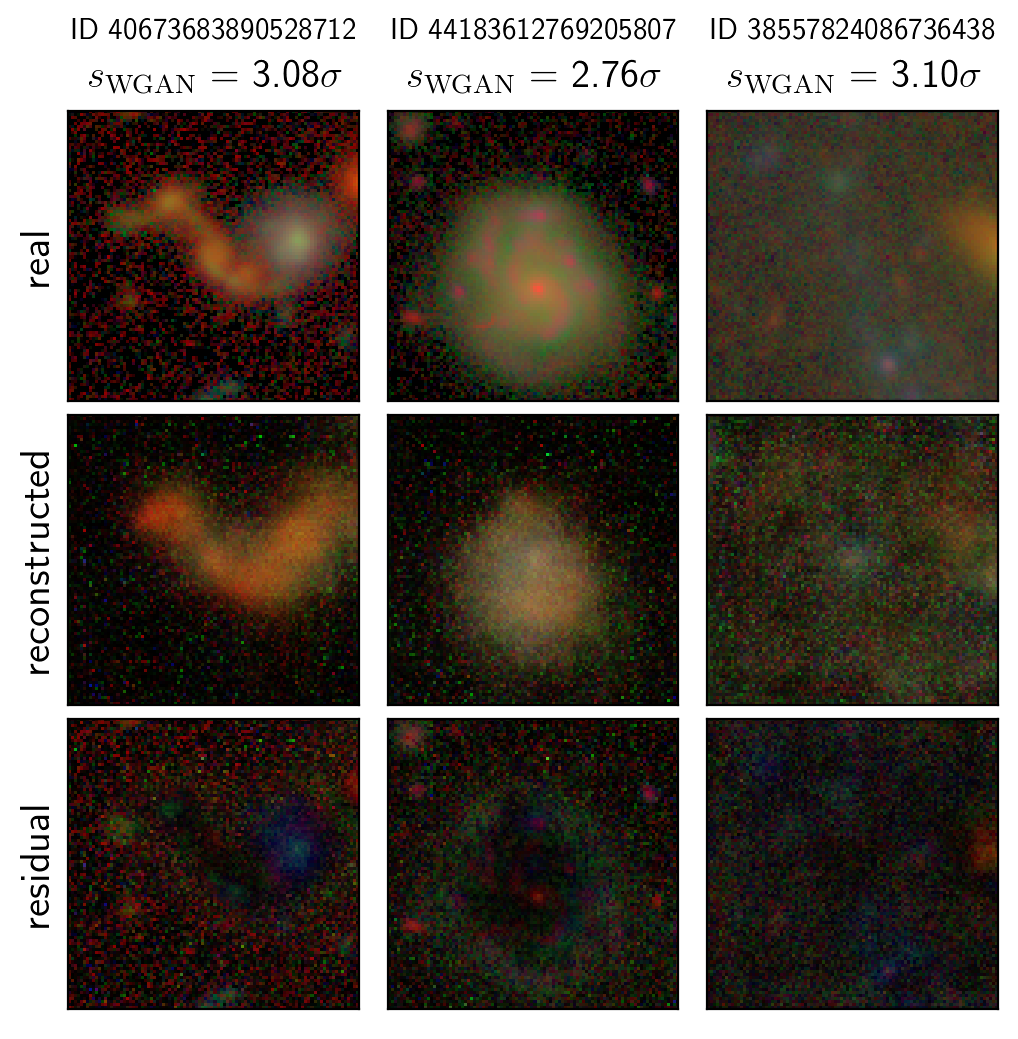}  
  \caption{}
  \label{fig:recon_3sig}
\end{subfigure}
\hfill
\begin{subfigure}{.325\textwidth}
  \centering
  \includegraphics[width=1\linewidth]{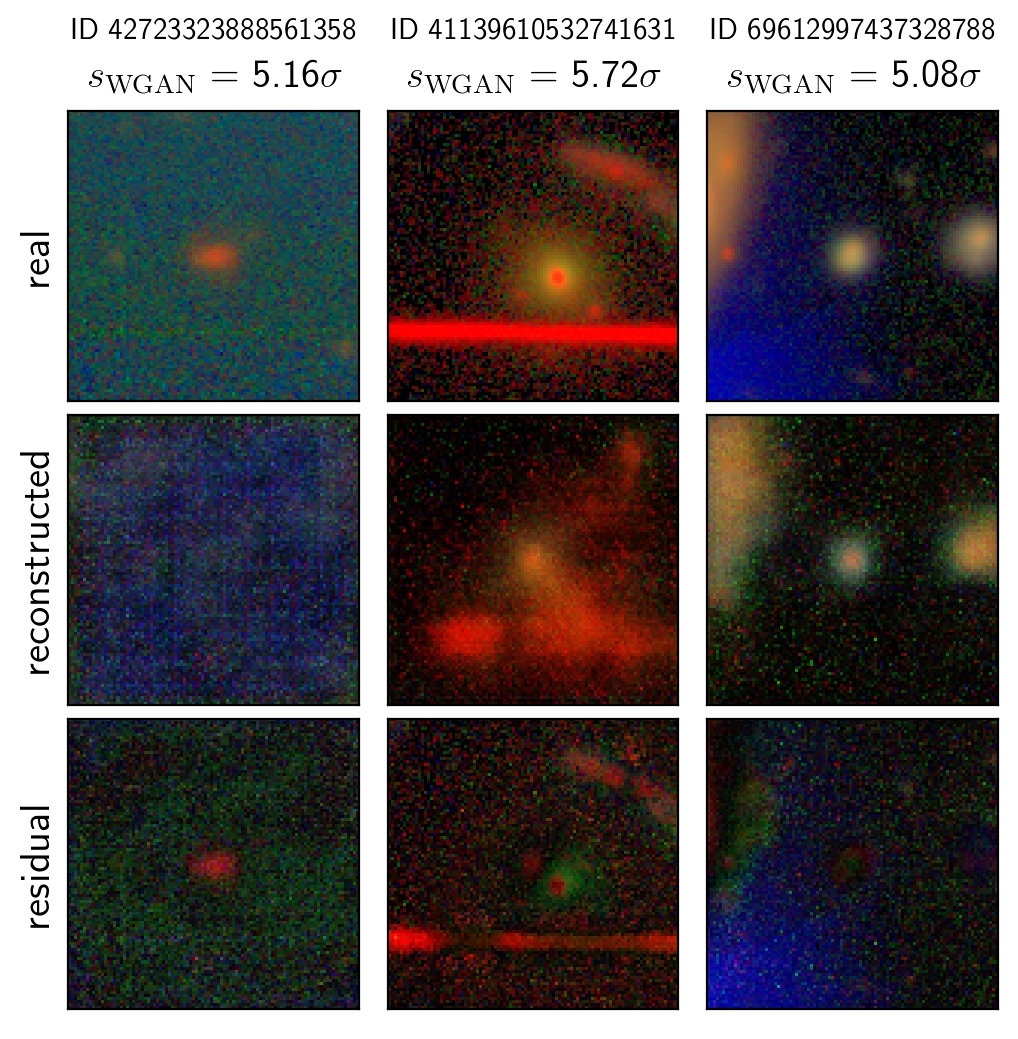}
  \caption{}
  \label{fig:recon_5sig}
\end{subfigure}
\caption{The results of WGAN image reconstruction and anomaly score assignment. The top row of each panel shows the original image, the second row shows the best WGAN reconstruction, and the bottom row shows the residual between the two. The assigned anomaly score is shown at the top of each column. The images in each panel are random samples of images in the following ranges of anomaly score: (a) significantly below the mean, (b) around $3\sigma$ above the mean, (c) greater than $5\sigma$ above the mean. It is clear that higher anomaly scores are indicative of poorer WGAN reconstructions and hence larger residuals.}
\label{fig:recon}
\end{figure*}

\begin{figure*}
\begin{subfigure}{.325\textwidth}
  \centering
  \includegraphics[width=1\linewidth]{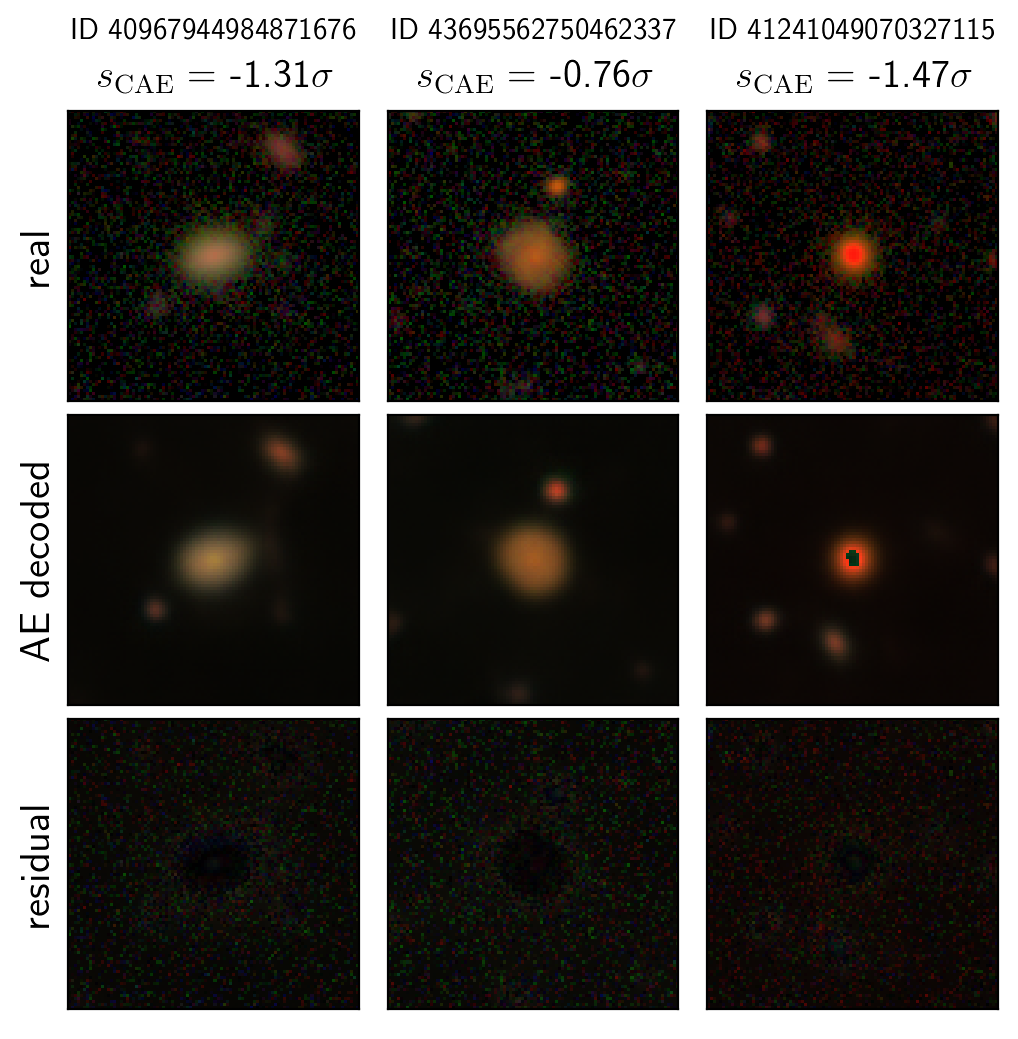}  
  \caption{}
  \label{fig:recon_ae_neg}
\end{subfigure}
\hfill
\begin{subfigure}{.325\textwidth}
  \centering
  \includegraphics[width=1\linewidth]{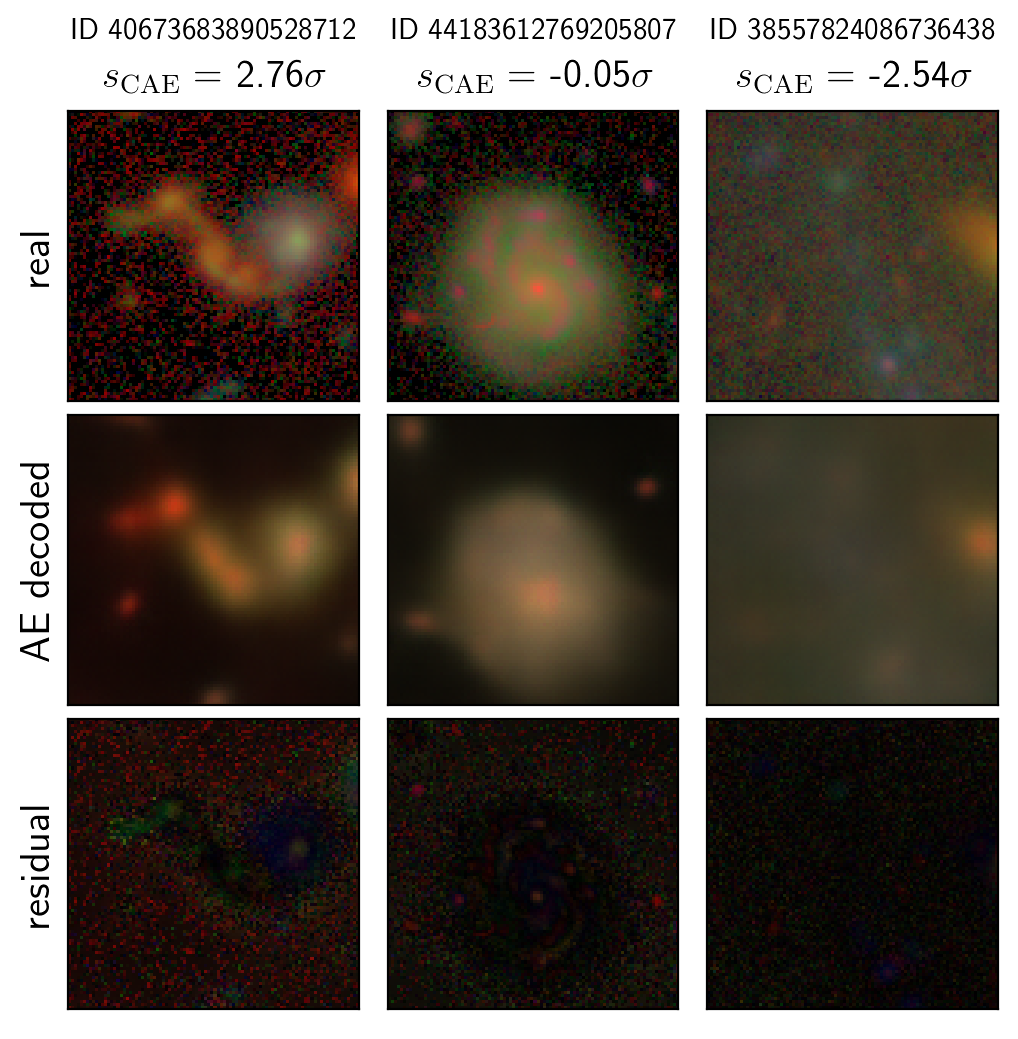}  
  \caption{}
  \label{fig:recon_ae_3sig}
\end{subfigure}
\hfill
\begin{subfigure}{.325\textwidth}
  \centering
  \includegraphics[width=1\linewidth]{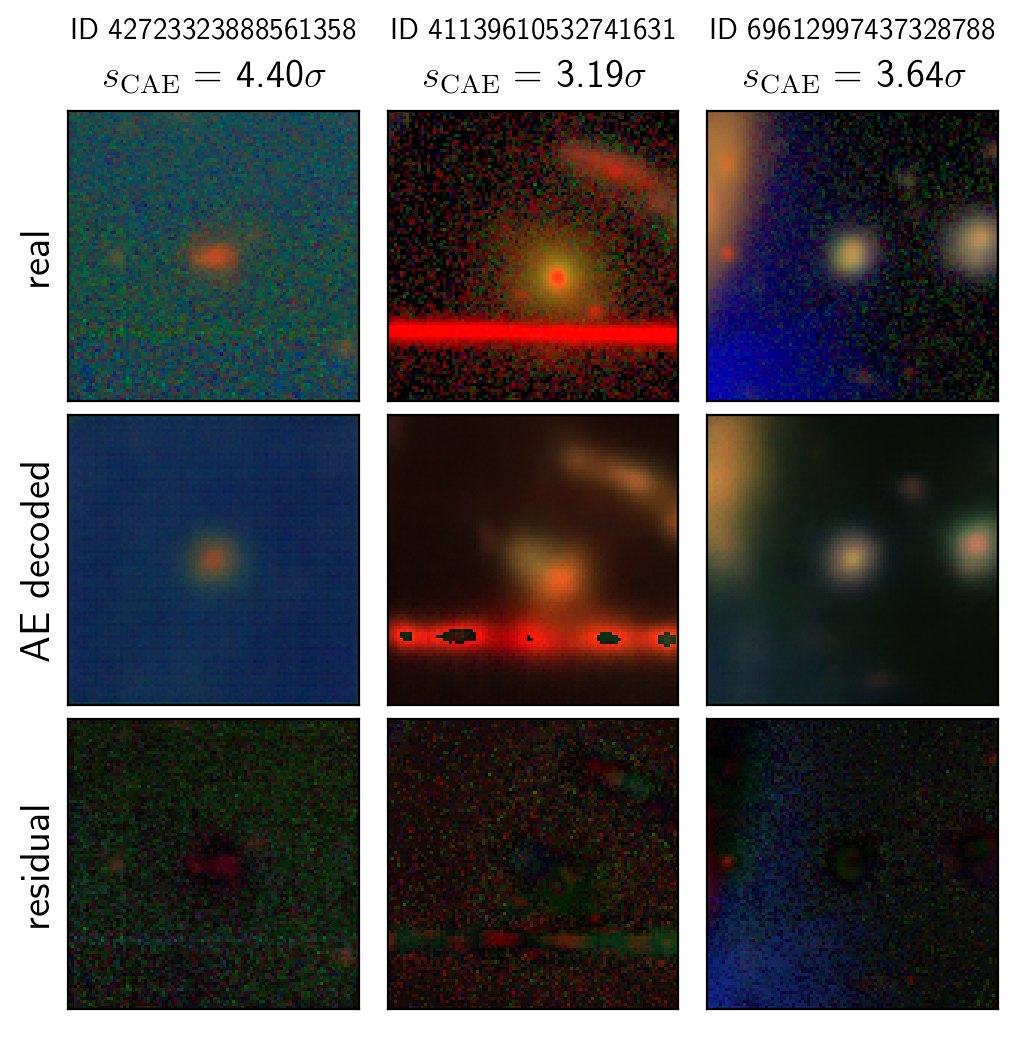}
  \caption{}
  \label{fig:recon_ae_5sig}
\end{subfigure}
\caption{\new{The results of CAE image reconstruction and anomaly score assignment. The rows are the same as in Figure~\ref{fig:recon}, and the same images are chosen. The assigned anomaly score is shown at the top of each column; note that the images are grouped by WGAN score, so the CAE scores are not consistent with the groups.}}
\label{fig:recon_ae}
\end{figure*}

The basic procedure for anomaly detection involves setting the WGAN to generate its best reconstruction of each image.
The WGAN has learned the global distribution of the data and will be better at generating ``typical'' images; therefore, we expect that a poorly reconstructed image indicates that an object is anomalous with respect to the rest of the sample.
This approach requires a quantification of how anomalous an object is and an inverse mapping from images to the WGAN's latent space.

To determine how well the WGAN reconstructs a given image and thus how anomalous it is, we set the trained network (with weights fixed after training) to find its best reconstruction of each image.
This is defined as the generator image that minimizes a loss $L$ based on the residuals between the original image and reconstructed image, in both pixel-space and feature-space.
The generator residual $L_\mathrm{gen}$ enforces a visual similarity between the images, and is defined as the pixel-wise difference between the original image $x$ and the generator-reconstructed image $G(\bm{z})$:
\begin{equation}
L_\mathrm{gen} = \Sigma_i^P \left| x_i - [G(\bm{z})]_i \right|
\end{equation}
where $i$ indexes the $P=96\times96\times3$ pixels in the image and its corresponding reconstruction.
We also use the discriminator to perform feature-matching to capture the similarity between the features of the reconstruction and the original.
The discriminator residual $L_\mathrm{disc}$ is calculated by considering the representation $d$ from the last convolutional layer, which has dimension $6\times6\times512$, and taking the difference between this representation for the real and reconstructed images:
\begin{equation}
L_\mathrm{disc} = \Sigma_j^F \left| d(\bm{x})_j - d(G(\bm{z}))_j \right|
\end{equation}
where $j$ indexes the $F$ discriminator features.
The total loss then $L_\mathrm{tot} = (1-\lambda_\mathrm{anom}) \, L_\mathrm{gen} \, + \, \lambda_\mathrm{anom} \, L_\mathrm{disc}$, where $\lambda_\mathrm{anom}$ is a weighting hyperparameter which we tune.
We choose $\lambda_\mathrm{anom}=0.3$ to balance the typical variation in raw scores between generator and discriminator residuals.
Varying this parameter results in slightly different samples of high-scoring anomalies; the importance of the generator vs. discriminator scores is investigated in Section~\ref{sec:sanom_dist}.

The inverse mapping from image to latent-space vector is typically performed with a straightforward optimization, starting from a random draw from the WGAN's latent space and optimizing to find the latent-space vector that minimizes $L$ (e.g. \citealt{Schlegl2017}).
However, this is a time-limiting step for large samples, so we propose an improvement: we first train an encoder, a standard convolutional network, on the entire training sample to make a first approximation of the latent-space vector.
This encoder simply provides a better initial guess of the latent-space location and does not significantly affect the final reconstruction.
We then start from the encoder approximation and, for each image individually, perform a basic minimization of $L$, optimizing for 10 iterations (though the score usually converges before this).
This loss value at the final iteration, $L_\mathrm{tot}^\mathrm{final}$, quantifies the degree of anomaly of the image, so we assign this to be the image's anomaly score, $s_\mathrm{WGAN} = L_\mathrm{tot}^\mathrm{final}$.
We also preserve the information about the generator and discriminator residuals, assigning associated scores $s_\mathrm{gen} = L_\mathrm{gen}^\mathrm{final}$ and $s_\mathrm{disc} = L_\mathrm{disc}^\mathrm{final}$.
Higher anomaly scores indicate more anomalous objects, while lower scores indicate objects better modeled by the WGAN; the scores are relative and meaningful only with respect to the rest of the sample.

The result of this process is shown in Figure \ref{fig:recon}.
We can see that the WGAN is able to generate realistic images; for compact objects with standard colors, it constructs images nearly identical to the original, and assigns the objects low anomaly scores (Figure \ref{fig:recon_neg}).
The model is more challenged to generate objects with rare features or colors, as in the images with scores around $3\sigma$ above the mean shown in Figure \ref{fig:recon_3sig}.
Finally, objects with optical artifacts, such as satellite streaks or contamination from nearby bright stars, have extremely high anomaly scores, as the WGAN struggles to reconstruct them (Figure \ref{fig:recon_5sig}).

\subsection{Dimensionality Reduction with a Convolutional Autoencoder}
\label{sec:cae}

A general problem with anomaly detection is to distinguish potentially interesting objects from trivial data issues.  
We propose here a new approach based on Convolutional Autoencoders (CAEs) to post-process and explore identified anomalies.

We expect the residual images, the difference between the real and reconstructed images, to contain information about why the WGAN marked an object as anomalous.
(We use the absolute difference because we are restricted to positive pixel values on the RBG scale, though this does lose potentially useful information.)
However, the pixel space is very high-dimensional, and contains information less relevant to the anomalous features we are interested in, such as background noise.
We employ a CAE to reduce the dimensionality of the data and isolate the relevant information.

We train a straightforward CAE to map the pixels of the residual images to a 64-dimensional vector.
The CAE has 4 encoding and 4 decoding layers, and uses a standard MSE loss between the true and reconstructed image.
We train the CAE in batches of 30 images, and stop the training when the loss stops improving, freezing the network at 30,000 iterations.
We then use the CAE to encode each of the images into a 64-dimensional vector representation.
We confirm that the CAE maps the encoded vectors to images that are reasonable reconstructions of the originals.
This autoencoding step allows us to extract the information most relevant to the anomalous features of the image, and perform further characterization to distinguish interesting anomalies; we demonstrate this in Section~\ref{sec:cae-umap}.
As a comparison, we also train an identically constructed CAE on the pixels of the real images; the results of this are also shown in Section~\ref{sec:cae-umap}.

\subsection{Anomaly Detection with a Convolutional Autoencoder as a Benchmark}
\label{sec:cae_bench}

\new{We compare our WGAN anomaly detection approach to a simpler method using a straightforward convolutional autoencoder.
We use the trained CAE described in Section~\ref{sec:cae}.
In that section we used the CAE for the purpose of dimensionality reduction in order to characterize the WGAN-detected anomalies, and we emphasize that that application is independent from this use of the CAE as a benchmark for anomaly detection.
For this comparison, we obtain the latent space representation from the trained CAE for each of the nearly one million HSC objects, and apply the trained decoder to get the CAE-decoded image.
We then compute a CAE anomaly score $s_\mathrm{CAE}$ by computing the residual image between the original and CAE-decoded image, and then taking the sum over the residual pixel values.
This is analogous to how we compute the generator score as described in Section~\ref{sec:sanom_assignment}.}

\new{We demonstrate the CAE approach in Figure~\ref{fig:recon_ae}.
We use the same sample of images as for the WGAN approach in Figure~\ref{fig:recon} for a direct comparison (note that the groupings are now misaligned with the CAE anomaly scores, as they are grouped by WGAN score).
The CAE decoded images are generally good reproductions of the original images, but the CAE also struggles for images that are outliers with respect to the full training set, as can be seen in Figure~\ref{fig:recon_ae_5sig}.
The CAE also produces very smooth images that lack the noise properties of the real data, while the WGAN is generally good at reproducing this noise.
We use the CAE anomaly score as a benchmark, as it is an established approach that is simpler than our WGAN method, but has some similar properties.}

\section{Results}
\label{sec:results}

\subsection{Anomaly Score Distribution}
\label{sec:sanom_dist}

\begin{figure*}
    \centering
    \includegraphics[width=\textwidth]{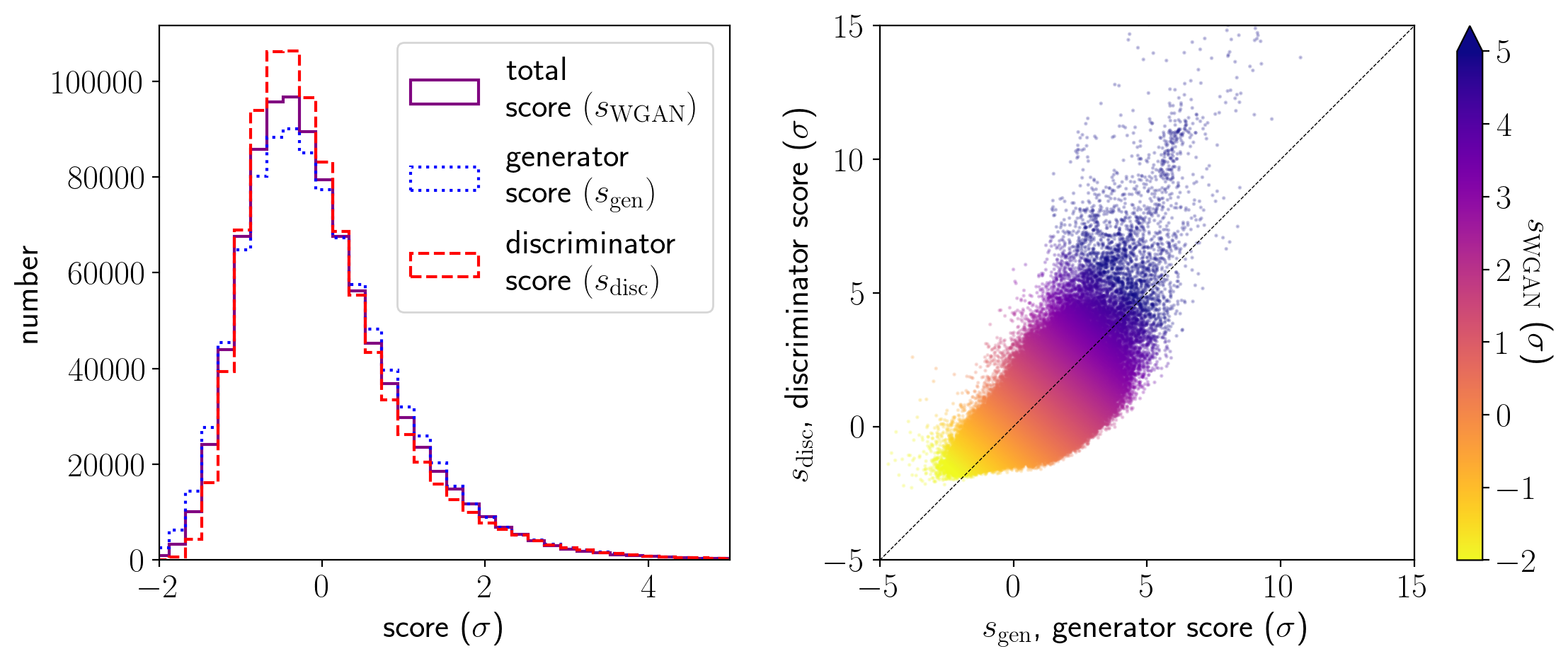}
    \caption{Left: The distribution of anomaly scores for the $\sim$940,000 objects in our sample, for the score based on the generator pixel-wise residual $\s{gen}$ (blue dotted), the score based on the discriminator feature-matching residual $\s{disc}$ (red dashed), and the combined total score $\s{WGAN}$ (purple solid). We normalize each by their mean and standard deviation $\sigma$, so the scores shown are in terms of $\sigma$ away from the mean of the given distribution. We do not show the long tails of these for clarity; a few objects extend up to $\s{gen}=25.5\sigma$ and $\s{disc}=35.2\sigma$, and down to $\s{gen}=-4.7\sigma$ and $\s{disc}=-2.3\sigma$.
    Right: The distribution of discriminator vs. generator scores, color-coded by total anomaly score. The dashed line shows where $\s{gen}=\s{disc}$ in terms of their $\sigma$ away from the mean.}
    \label{fig:dist}
\end{figure*}

\begin{figure*}
    \centering
    \includegraphics[width=\textwidth]{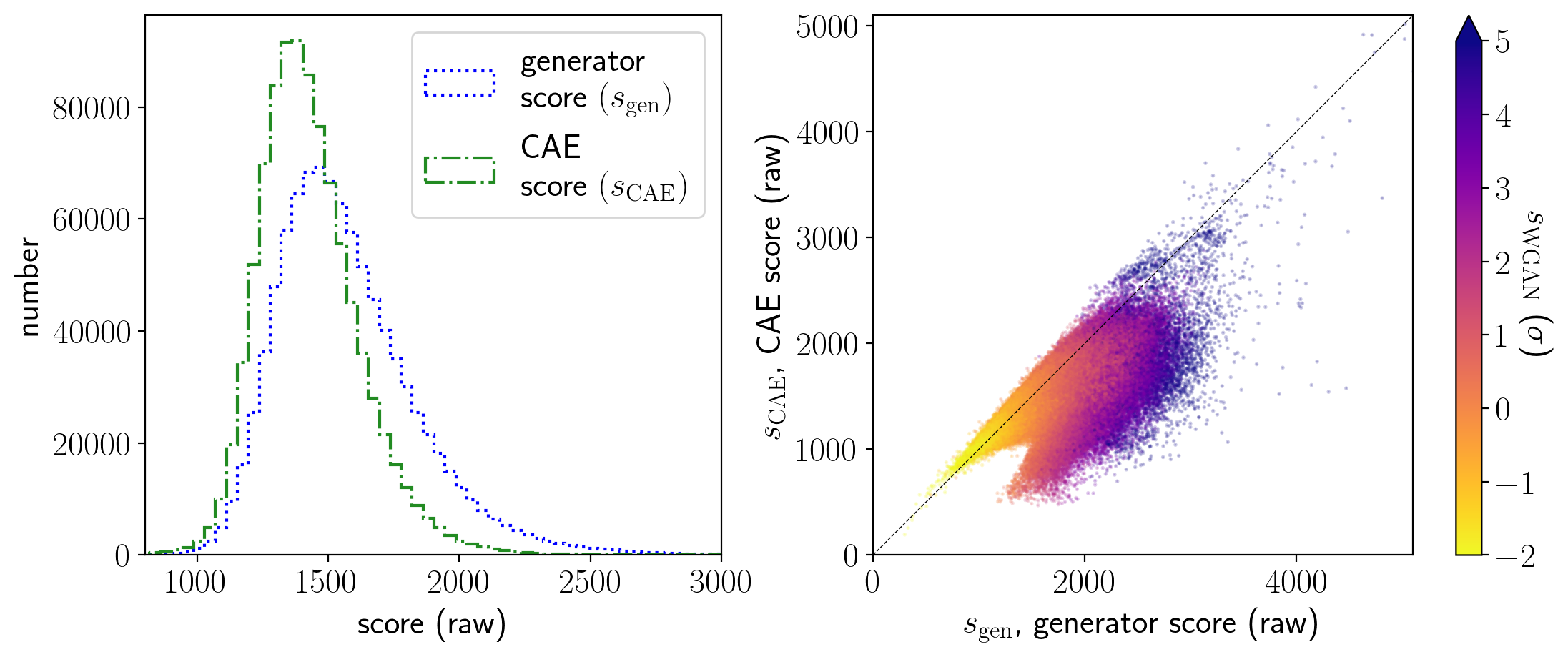}
    \caption{\new{Left: The distribution of anomaly scores for the $\sim$940,000 objects in our sample, for the score based on the generator pixel-wise residual $\s{gen}$ (blue dotted), and the score based on the CAE pixel-wise residual (green dot-dashed). We show the raw scores, as these are directly comparable. As in Figure~\ref{fig:dist}, we do not show the long high-score tails for clarity.
    Right: The distribution of generator vs. CAE scores, color-coded by total WGAN anomaly score. The dashed line shows where $\s{gen}=\s{CAE}$.}}
    \label{fig:dist_ae}
\end{figure*}

We compute anomaly scores for each of the $\sim$940,000 objects in our sample, using both a generator score $\s{gen}$ and a discriminator score $\s{disc}$ as described in Section~\ref{sec:sanom_assignment}.
These raw scores have very different ranges due to their definitions based on image pixel differences or feature value differences, so to compare them we show the scores by their number of standard deviations from the mean. 

These distributions are shown in the left panel of Figure \ref{fig:dist}.
Recall that higher anomaly scores indicate more anomalous objects, while lower scores indicate objects that are more well-modeled by the WGAN.
We find that the distribution is skewed towards higher scores, which is expected: most typical objects are reconstructed well by the WGAN so have similar scores, while there are more ways to be anomalous than to be typical, resulting in a wider range of scores.
The high-score tail extends out to an object with $\s{disc}=35.2\sigma$, and the low-score tail extends to an object with $\s{gen}=-4.7\sigma$ (we only show the bulk of the distribution for clarity).

The right panel of Figure \ref{fig:dist} shows $\s{disc}$ vs.  $\s{gen}$ for each of the images.
As expected, we see that most objects have relatively similar generator and discriminator scores, indicating that the generator and discriminator generally agree on the degree of anomaly of the image.
That said, the distribution has significant scatter, so the pixel-residuals and feature-residuals may be picking up on different indications of anomalousness.
At high scores, there is a skew towards higher discriminator scores: anomalous objects are more likely to have a high discriminator score compared to generator score.
However, this only applies to very high-scoring objects above $\sim5\sigma$, and is likely a result of how we compute the raw scores---there is a maximum to the generator residual due to the pixel values, while discriminator scores are essentially unbounded---so we do not subscribe great significance to this.
These distributions also reflect the fact that the scores are not Gaussian distributed, but rather have a long one-sided tail towards high anomaly scores.

\new{As a comparison, we also compute the CAE anomaly scores for each of the images as described in Section~\ref{sec:cae_bench}.
These are based on the residual pixel values between the real image and the CAE decoded image, and so they are directly comparable to the generator scores.
We show the raw score distribution for both of these in the left panel of Figure~\ref{fig:dist_ae}.
The CAE scores are lower on average than the generator scores, meaning the reconstructions are generally more similar to the original image.
The right panel shows the correspondence between the generator and CAE scores for each image, and we see that the CAE scores are typically lower.
These points are color-coded by the total WGAN score; we see that there is also not much of a correlation between CAE score and WGAN score, meaning that the discriminator component of the WGAN score corresponds more with the generator score than the CAE score.}

\new{Although the CAE scores tend to be lower than the WGAN scores, we hypothesize that this is largely due to noise properties.
Looking at Figures~\ref{fig:recon} and \ref{fig:recon_ae}, the CAE reconstructions are much smoother and less noisy than the WGAN reconstructions.
The residuals for the CAE then contain the noise of the real image.
On the other hand, the WGAN reconstructions attempt to generate images with similar noise properties as the original, as the WGAN has learned the typical noise level of the training data.
However, it will not place the noise in exactly the same pixels as in the real image (given that we are limited in the optimization step to find the best reconstruction of the image in the WGAN's latent space).
The residuals for the WGAN then contain roughly double the noise values as those for the CAE, as they include the noise of the original and the reconstruction. 
This contributes significantly to the generator score, pushing the score distribution to higher values compared to the CAE scores.
If this difference is mainly due to noise, then it is not very relevant to anomaly detection; the differences in scores within each definition will be more meaningful.
We investigate this in the next section by looking at high-anomaly samples based on each score definition.
We also note that in some cases the CAE does seem to produce a better reconstruction than the WGAN, but we show in the next section that even so, the WGAN discriminator score is a more useful metric for identifying interesting anomalies.}

\subsection{High-Anomaly Sample Selection}
\label{sec:anom_sample}

\begin{figure*} 
\begin{subfigure}{0.33\textwidth}
  \centering
  \includegraphics[width=0.95\linewidth]{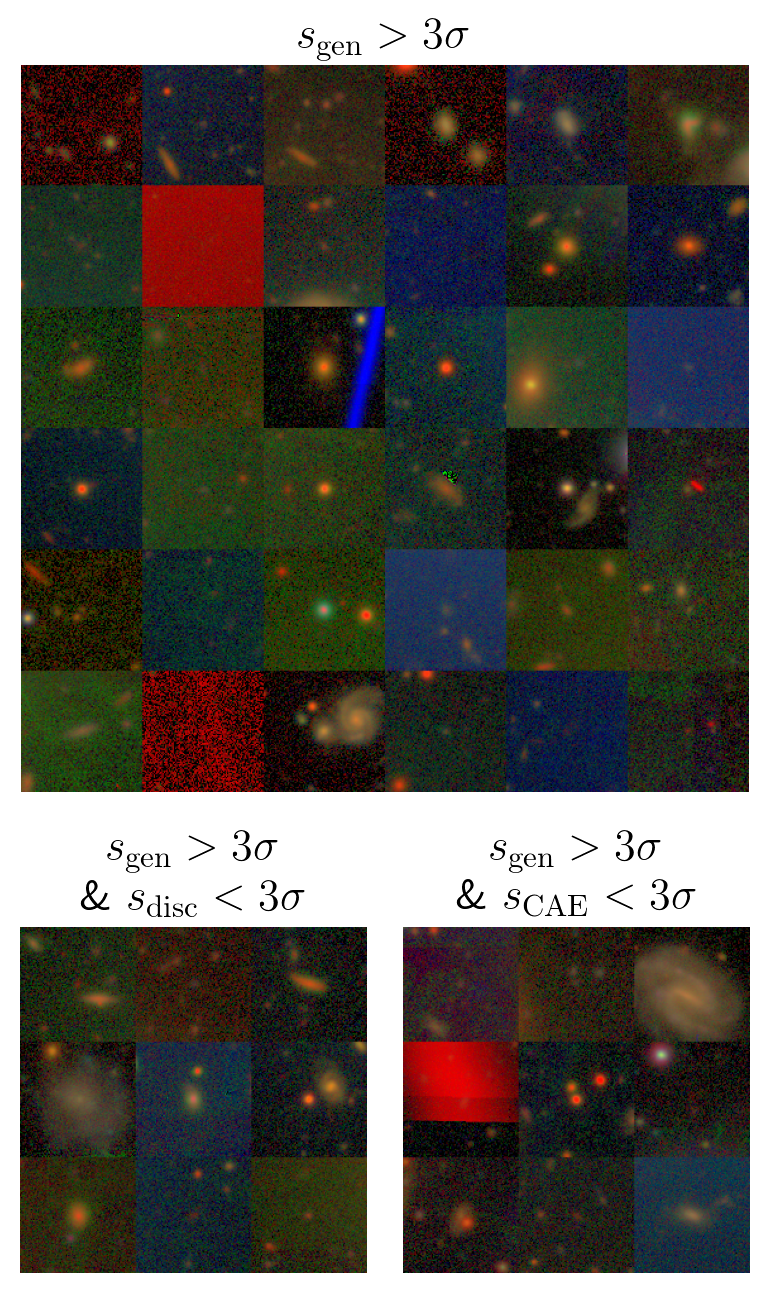}
  \caption{}
  \label{fig:score_effect_gen}
\end{subfigure}
\begin{subfigure}{0.33\textwidth}
  \centering
  \includegraphics[width=0.95\linewidth]{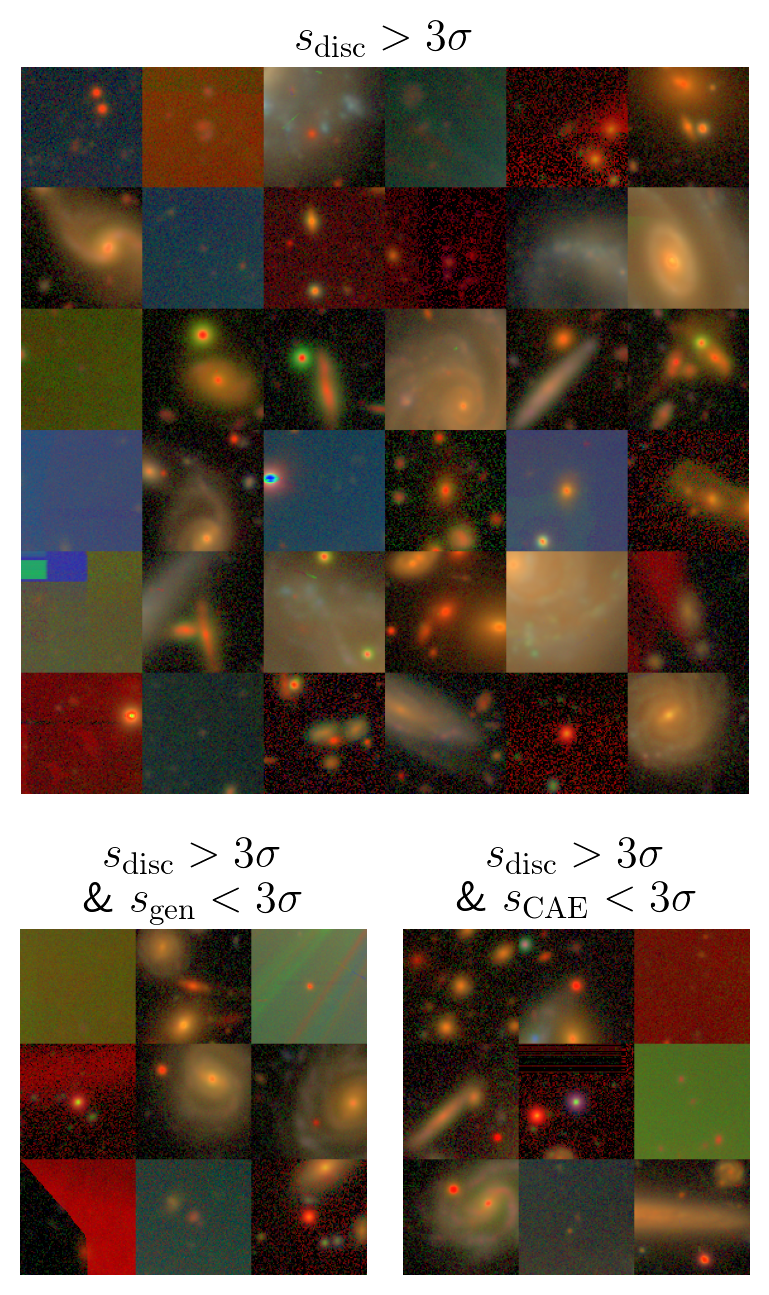}
  \caption{}
  \label{fig:score_effect_disc}
\end{subfigure}
\begin{subfigure}{0.33\textwidth}
  \centering
  \includegraphics[width=0.95\linewidth]{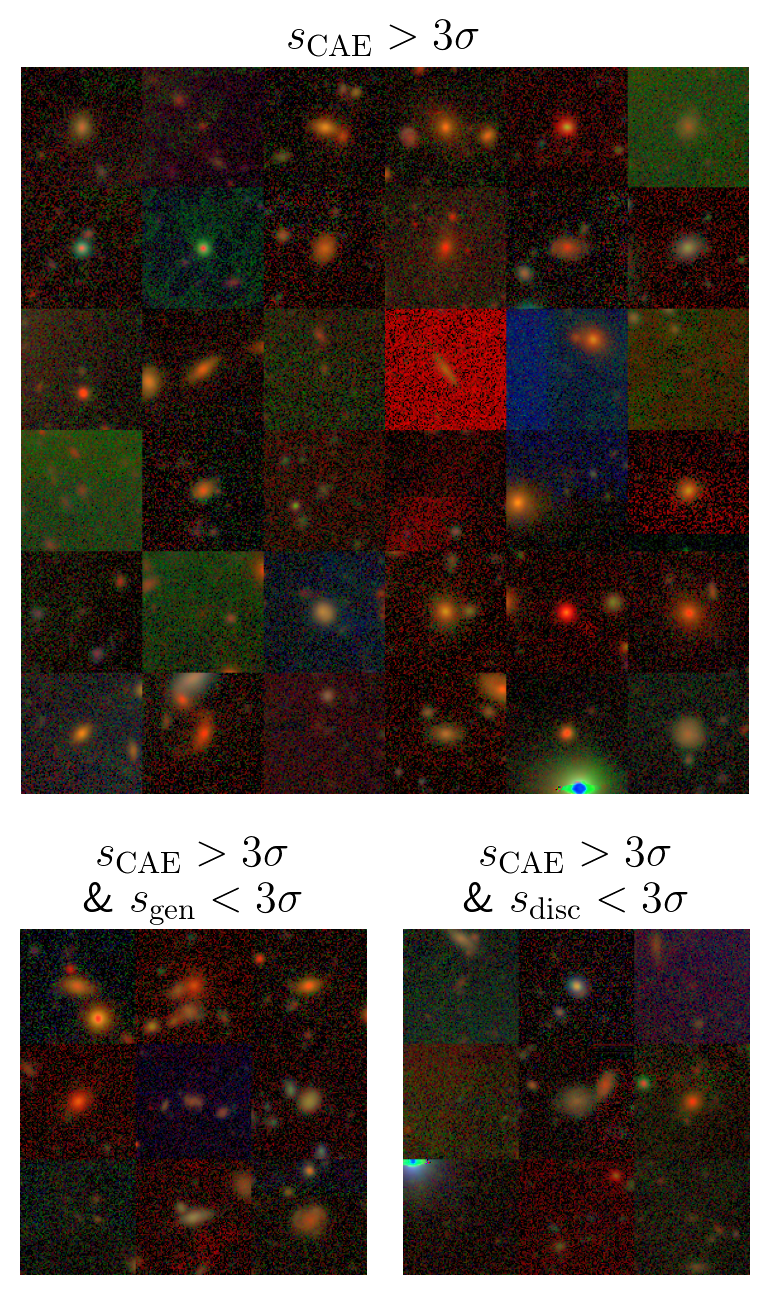} 
  \caption{}
  \label{fig:score_effect_ae}
\end{subfigure}

\caption{Comparison between image samples selected using various score definitions. The top panel of (a) shows a random sample with a 3$\sigma$ cut on the combined anomaly score $\s{gen}$. The lower left panel shows a sample of images that this score definition would include but $\s{disc}>3\sigma$ would miss; the lower right panel of (a) similarly shows images missed by $\s{CAE}>3\sigma$. \new{Panels (b) and (c) show the same thing for $\s{disc}$ and $\s{CAE}$}. The $\s{disc}>3\sigma$ selection shows the highest proportion of interesting anomalies, and it selects interesting images the other scores would miss; we choose this selection criterion for our high-anomaly sample for further characterization.}
\label{fig:score_effect}
\end{figure*}

We investigate the role of the generator and discriminator scores by looking at selections of high-scoring anomalies with different score definitions.
We also compare these to selections based on the CAE score.
Figure~\ref{fig:score_effect} shows a comparison of random samples of these definitions, based on a $3\sigma$ cutoff.
The larger samples in the top row of all panels show a selection based on $\s{gen}$, $\s{disc}$, \new{and $\s{CAE}$}.
It is clear that all of these high-scoring samples are anomalous with respect to the full sample (Figure~\ref{fig:real}), showing objects with interesting features and properties, as well as noise-dominated images and those with optical artifacts.
We can see that the generator score sample (Figure~\ref{fig:score_effect_gen}) contains many empty images and some containing optical artifacts; the majority of the images are clearly scientifically uninteresting. 
The discriminator score sample (Figure~\ref{fig:score_effect_disc}) also has some noisy and saturated images, but it contains a significantly higher proportion of interesting-looking images, which we broadly consider as those that contain actual galaxies, and galaxies that are not just typical-colored, compact sources.
It also contains some objects of even higher interest, including extended galaxies with unusual structure, and compact objects with extreme colors.
\new{Looking at the CAE score sample (Figure~\ref{fig:score_effect_ae}), we see that the high CAE scores are picking up many images with high, grainy noise, as well as some optical artifacts.}

For each selection, we can ask which images are captured that other score definitions would miss with the $3\sigma$ cut; this is shown in the bottom row of each panel.
We can see that the generator score selects mostly noisy images that the other definitions miss (bottom panels of Figure~\ref{fig:score_effect_gen}). 
The discriminator score definition, on the other hand, would capture a decent number interesting objects that are missed by both the generator and CAE definitions (bottom panels of Figure~\ref{fig:score_effect_disc}).
\new{The high-scoring CAE images that the other definitions miss (bottom panels of Figure~\ref{fig:score_effect_ae}) are, once again, mainly noise and otherwise uninteresting images.}

This analysis suggests that the generator score, which is based on the pixel-residual, is more attuned to selecting anomalous images due to noise or other image corruption issues, which tend to affect most of the pixels in the image \new{and lead to larger absolute differences between the real and reconstructed images}.
On the other hand, the discriminator selects images that are anomalous in feature-space, which is more flexible in selecting various types of anomalies including those that are more localized in pixel space.
This results in a set of images that have more potential to be scientifically interesting.
\new{We also investigate using the combined score for selecting this high-anomaly sample, and find that it effectively results in a mix of the generator and discriminator high-scoring samples; as we showed in Figure~\ref{fig:score_effect} that the generator score selects very few interesting objects missed by the discriminator score, we choose to not incorporate the generator score at all and just use the discriminator score.}

\new{We hypothesize that the WGAN is better suited to anomaly detection compared to the CAE due to several factors.
The first is that the WGAN is specifically intended to learn the true distribution of the input data.
Thus we should be able to better identify the objects that are outliers of that distribution.
On the other hand, the CAE is much simpler, just trained to compress and reproduce the images in a lower dimension.
Second, CAEs tend to smooth the data, as we see in our decoded images; this will wash out subtle anomalies, and in astrophysics we often care about these subtleties.
Finally, the WGAN's discriminator is particularly suited to anomaly detection, as it is trained to detect ``realistic'' images, which will look more like the bulk of the input data.
We can use the discriminator feature-space as a natural anomaly quantification, and we see that indeed it finds more interesting anomalies, while the CAE has no such quantity.}

We thus select our final high-anomaly sample with a cut on discriminator score, $\s{disc}>3\sigma$, as in the upper panel of Figure~\ref{fig:score_effect_disc}, producing a sample with 13,477 images, representing 1.4\% of the total sample. 
We also show $\s{disc}$ as the relevant anomaly score in the rest of our analysis given its stronger relation with potentially interesting anomalies.
We note that the choice of using only $\s{disc}$ for our anomaly score is different than previous work, such as \cite{Schlegl2017}, \cite{Zenati2018a}, and \cite{Margalef-Bentabol2020}, which all use an anomaly score that combines the generator and discriminator losses.
Our choice may indeed miss some anomalies, but we have shown in Figure~\ref{fig:score_effect} that the choice of a $\s{disc}>3\sigma$ results in missing the least interesting objects.
We strive for purity over completeness in our high-anomaly sample, as we aim to identify scientifically interesting anomalies, which is easier with less contamination from noisy images.
We provide the full score information in our released catalog so subsequent studies may use a different selection if they so choose.

\subsection{Correlation with HSC Catalog Information}

\begin{figure*}
    \centering
    \includegraphics[width=0.8\textwidth]{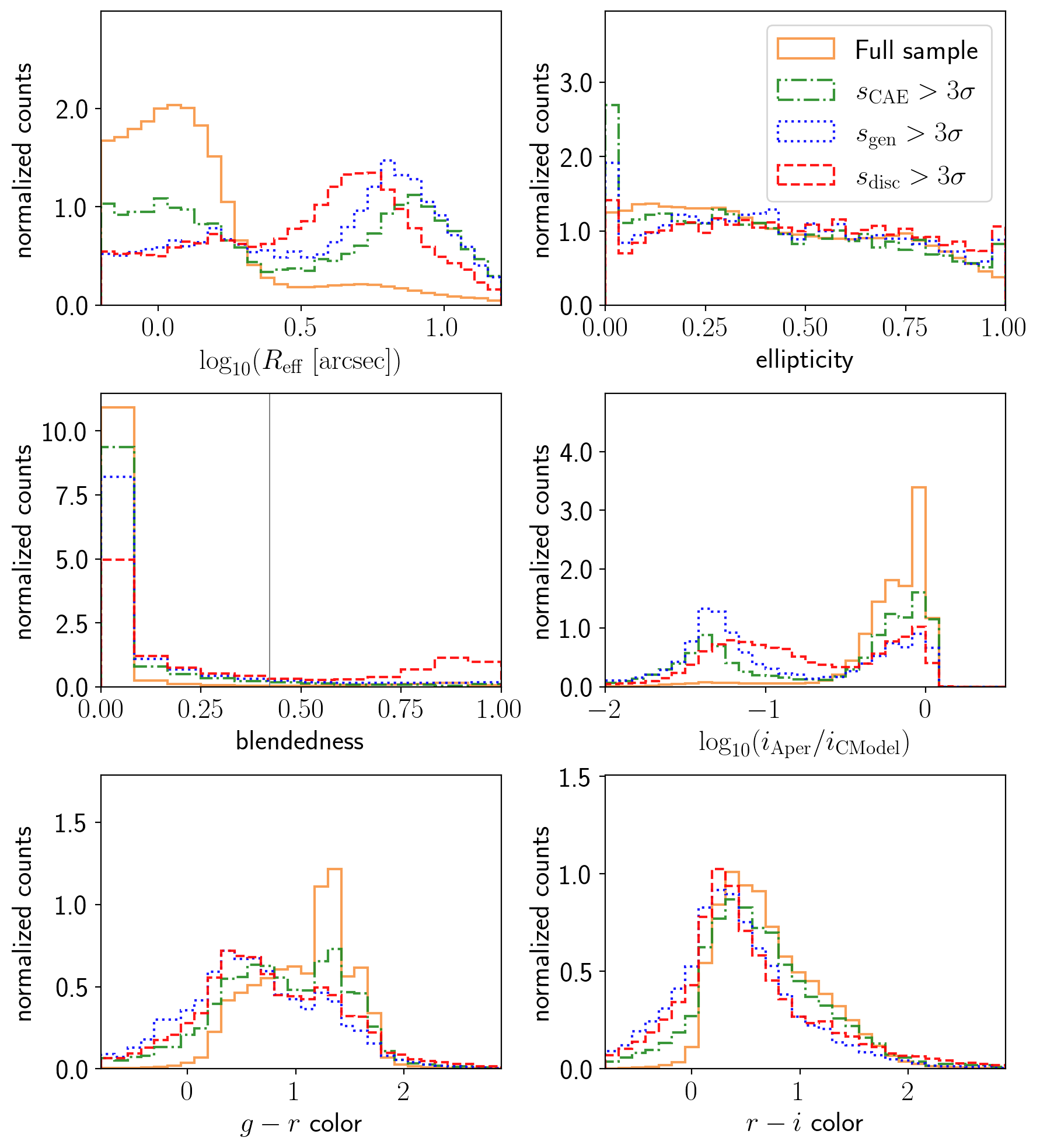}  
    \caption{The distribution of catalog properties from the HSC pipeline for various score definitions. We compare the distribution of the full sample (solid orange) to those of high-anomaly samples selected by \new{$\s{CAE}$ (dot-dashed green)}, $\s{disc}$ (dashed red), and $\s{gen}$ (dotted blue). The properties we show are the effective radius $R_\mathrm{eff}$, ellipticity, blendedness, ratio of aperature to CModel flux $i_\mathrm{aper}/i_\mathrm{CModel}$, $g-r$ color, and $r-i$ color. The grey line in the blendedness panel indicates the typical threshold for filtering out highly blended objects.}
    \label{fig:hsc_hist}
\end{figure*}

In order to further explore the results of our anomaly detection process, we compare the objects in various samples with derived properties from the HSC catalog.
This acts as a validation step to understand the information that our anomaly detection approach might be using to make its assignments, \new{as well as to} determine if it is picking up on information beyond that in the catalog.
Figure~\ref{fig:hsc_hist} shows the normalized distributions of selected galaxy properties for the full sample and the $3\sigma$-anomaly samples, selected with each of the score definitions discussed previously.

We first look at the spatial extendedness of the object.
We use the effective radius $R_\mathrm{eff}$ of the exponential component from the best-fit CModel result. 
It indicates the intrinsic extendedness of an object after taking PSF convolution into account. 
Among the different size measurements provided in HSC database, it is the most stable and robust measure of size.
Objects with $R_\mathrm{eff}<1''$ are very compact, while those with $R_\mathrm{eff}>5''$ are quite extended (recall that the sizes of cutouts are $15 \times 15$ arcsec).
Among the objects with $R_\mathrm{eff}>5''$, we notice a population of problematic objects that do not appear to be that extended and also have faint $r$- and $g$-band magnitude; we believe that these are due to bad deblending processes.
From the figure, we see that most of the objects in the full sample are compact, with a median of $R_\mathrm{eff}\sim0.5''$. 
In comparison, the generator and discriminator $3\sigma$ anomaly samples contain a majority of extended objects; the discriminator-selected sample has a median $R_\mathrm{eff}\sim2.5''$.
This shows that the WGAN tends to find high-$R_\mathrm{eff}$ objects to be more anomalous, as these high values encompass both bad deblends and actual extended galaxies.
The latter makes sense as extended objects are more resolved and can show more complex details, such as interesting galactic structure.
That said, these high-anomaly samples still contain significant numbers of compact objects, showing that the WGAN is able to detect interesting compact objects and is not simply flagging all extended objects as anomalous and all compact ones as boring.
\new{We note some small but interesting differences between the generator and discriminator samples:} the generator selects many objects with unrealistically high radii, while the discriminator tends towards intermediate-extended objects.
\new{We also show the distribution of high-CAE score objects, and we find that it separates into two populations: one of compact objects like the bulk of the full sample, and the other of very large radius objects that are likely bad deblends.
This latter peak is very similar to (and even more extreme than) that feature in the generator sample distribution.}
This supports our finding that the generator \new{and the CAE} are attuned to optical artifacts which do not have meaningful radii, while the discriminator more often finds actual galaxies with interesting properties.

We next examine the ellipticity $e$ of objects using the exponential components of the CModel flux.
Here 0 means circular and 1 means highly elliptical.
We note that this is for both extended and compact objects, and ellipticity may be less meaningful for the latter.
In the full sample, objects skew towards small ellipticity.
While the full population has a higher fraction of objects in the $0.05 < e < 0.3$ bin, the anomaly samples show an excess of very low ellipticity objects at $e < 0.05$.
\new{Interestingly, the CAE sample has the highest excess, perhaps due to having a higher proportion of compact objects.}
The anomaly samples also tend to pick up more highly elliptical ($e>0.9$) objects. 
Such high ellipticity is rarely physical; it is often due to poor fits on corrupted images.

We look at the blendedness of the objects, which describes the contamination of one object by the light from other close objects. 
It is computed from the $i$-band flux fits, as described in \citep{Bosch2019}.
A blendedness of 0 indicates an isolated object, while a value near 1 indicates a very blended object. 
A very small number of objects are assigned unphysical negative blendedness scores; we do not show these, but they do contain a higher proportion of anomalies.
We see that in the full sample, for the vast majority of objects, their photometry is not affected by image blends.
For the anomaly samples, most images also have low blendedness values, showing that the anomaly detector is finding interesting isolated objects and not just identifying blends.
That said, the anomaly samples do also show a significantly higher fraction of highly blended objects compared to the full sample.
This is expected due to the low representation of blended objects in the data, as well as the correlation between objects with larger angular size and higher blendedness.
The grey line at $10^{-0.375}$ shows the cutoff suggested for eliminating blends \citep{Mandelbaum2018}; 4.6\% of all of the objects in our full sample are above this threshold, while 32\% of the objects in the high discriminator score sample are above it.
Interestingly, the discriminator selects for more of the highly blended objects than the generator \new{and CAE}, suggesting that many of these are actual overlapping objects and not just artifacts, as we know that the discriminator flags fewer artifacts as anomalous compared to the \new{other scores}.

The ratio between the aperture flux and CModel flux $\log_{10} (i_{\rm Aper}/i_{\rm CModel})$ is another interesting diagnosis of photometry. 
It has been used as an empirical indicator for problematic objects due to bad deblending processes.
Figure \ref{fig:hsc_hist} only shows the $i$-band, but the distribution is very similar for the other bands.
The aperture flux is computed in a small $2''$ aperture on images convolved with kernels matched to a common FWHM-1.2 arcsec PSF across all five bands, to homogenize the flux measurements across the bands; the CModel flux takes into account the emission from the entire image. 
Thus a low ratio means that the object is very large, or that there is a strong color gradient; it may also indicate that there is an issue with the CModel fit, for instance due to poor deblending.
We see that most objects in the full sample have a ratio near one, with a skew towards smaller ratios as expected from the flux definitions.
The $3\sigma$ anomaly samples show a markedly different distribution, with many objects having very small ratios.
This makes sense as we have seen that our anomaly detector selects for both extended objects and deblending errors.
\new{The generator-selected sample shows a large proportion of very small ratio objects: 56\% of the high generator score images have a $\log_{10} (i_{\rm Aper}/i_{\rm CModel}) < -1.0$, compared to 40\% of discriminator score images, 36\% of CAE score images, and 4.8\% of the full sample.
This aligns with the generator's tendency to flag optical artifacts that likely have poor deblends, and shows that this metric is particularly useful for understanding the types of images that our anomaly detectors are selecting.}

Finally, we look at the color distributions for both $g-r$ and $r-i$ color, based on the CModel flux fits.
We see that the high-anomaly samples are significantly bluer (lower $g-r$) than the full sample.
There is less of a difference for $r-i$ color, but the anomalies do skew towards bluer colors (lower $r-i$).
This may be because of the relationship between color and angular size: galaxies with larger $R_\mathrm{eff}$ tend to be bluer, and larger objects are disproportionately anomalous in our sample, as we have seen in the $R_\mathrm{eff}$ distribution.
In both cases, we also see a longer tail into the red end of the color distribution, indicating that the anomalous sample contains more extreme color images on both ends; this makes sense as both interesting objects and problematic photometry would exhibit extreme colors.
We see little difference in the distributions for the \new{discriminator and generator $3\sigma$ samples}, suggesting that they are capturing similar proportions of objects of each color.
\new{The CAE $3\sigma$ score distribution is more similar to the full distribution, as we have seen in the previous panels, indicating that it is not picking up on as many outlying images as the WGAN.}

We also note that there is a strong correlation between objects near bright stars and their anomaly score.
For our initial selection we performed a basic cut on objects near foreground stars, but this still left significant potentially contaminated regions.
Applying a more aggressive bright star mask, we see that 44.6\% of objects in our discriminator-defined high-anomaly sample fall within the mask, compared to only 8.5\% of objects in the full sample (this newer mask can be found at \url{https://hsc-release.mtk.nao.ac.jp/doc/index.php/bright-star-masks-2}).
For increased purity, we could apply this mask, though we would risk losing some potentially interesting anomalies; in fact, these regions are often excluded due to the amount of artifacts, but a filter based on anomaly score could make this data usable.
For this work we do not apply this mask but note that it may be useful depending on the downstream application.
\new{Additionally, future HSC data releases that are in preperation will have more sophisticated data reduction and filtering, and will allow anomaly detectors to better learn interesting anomalies.}

This comparison of the distributions of catalog properties elucidates the types of images that the WGAN generator and discriminator, \new{and in comparison the CAE}, find to be anomalous. 
\new{The high anomaly score samples generally have significantly different distributions of properties compared to the full sample}. 
\new{In particular, the objects with high anomaly scores tend to have more extreme properties, including} both potentially interesting and unphysical properties.
\new{We also observe that the CAE- and generator- selected high-scoring samples have relatively similar distributions to each other for many of the properties, more so than the discriminator-selected sample, confirming the tendency of the CAE and generator to select more for noisy images and optical artifacts.}
\new{Further,} we found that there is a significant number of objects with non-extreme catalog properties but that are found to be highly anomalous; this validates that the anomaly detectors are incorporating more or higher-order information in determining the degree of anomaly of an image, \new{so these anomaly samples could not be constructed with simple cuts on pipeline data}.
Our anomaly scores could be used in combination with the catalog to further filter out noisy and uninteresting images or to pick out specific types of anomalies.

\subsection{Autoencoder Results Visualized with UMAP Embedding}
\label{sec:cae-umap}

\begin{figure*} 
\begin{subfigure}{.48\textwidth}
  \centering
  \includegraphics[width=1\linewidth]{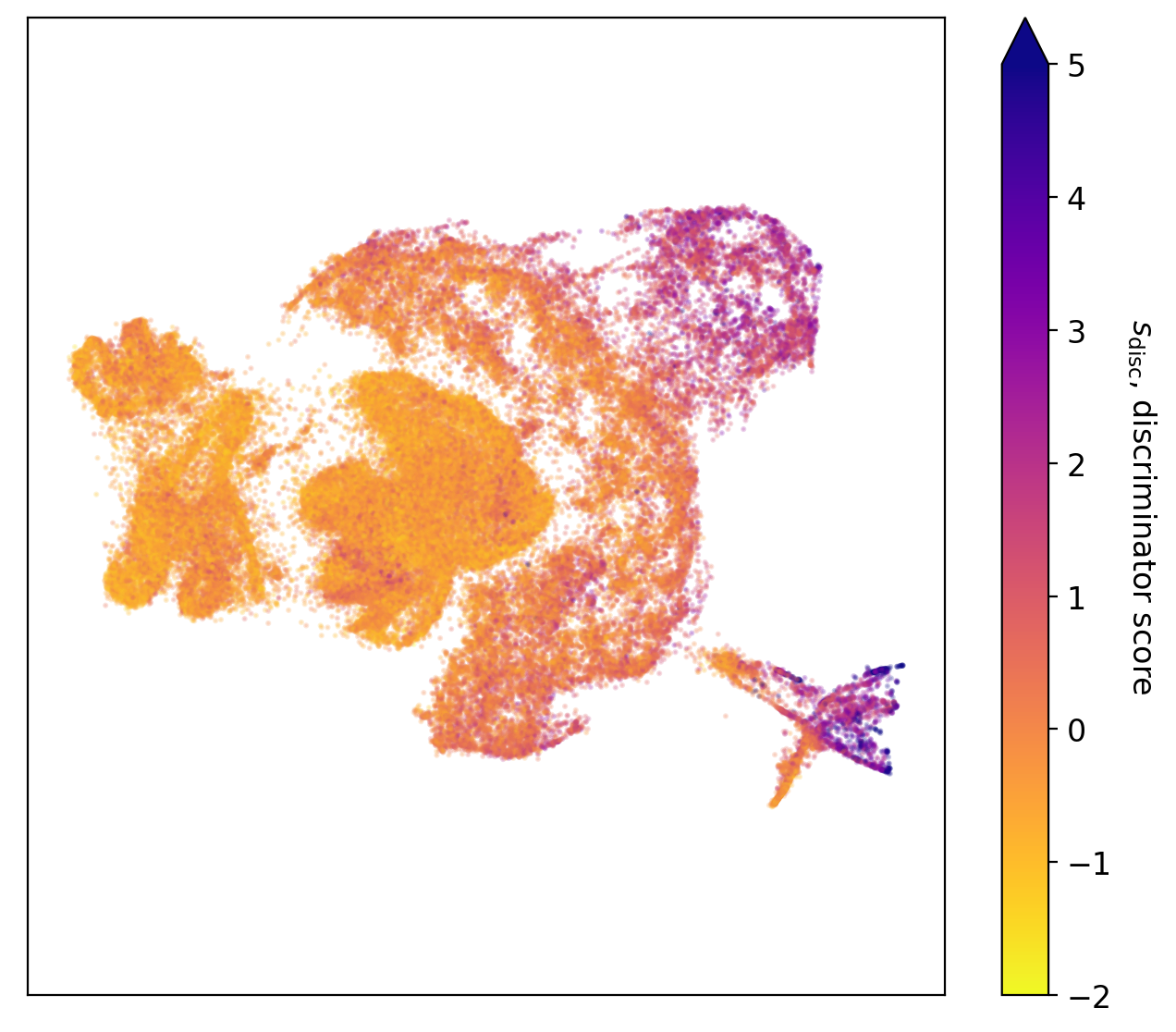}  
  \caption{Embedding with original image pixels.}
  \label{fig:umap_100k_reals}
\end{subfigure}
\begin{subfigure}{.48\textwidth}
  \centering
  \includegraphics[width=1\linewidth]{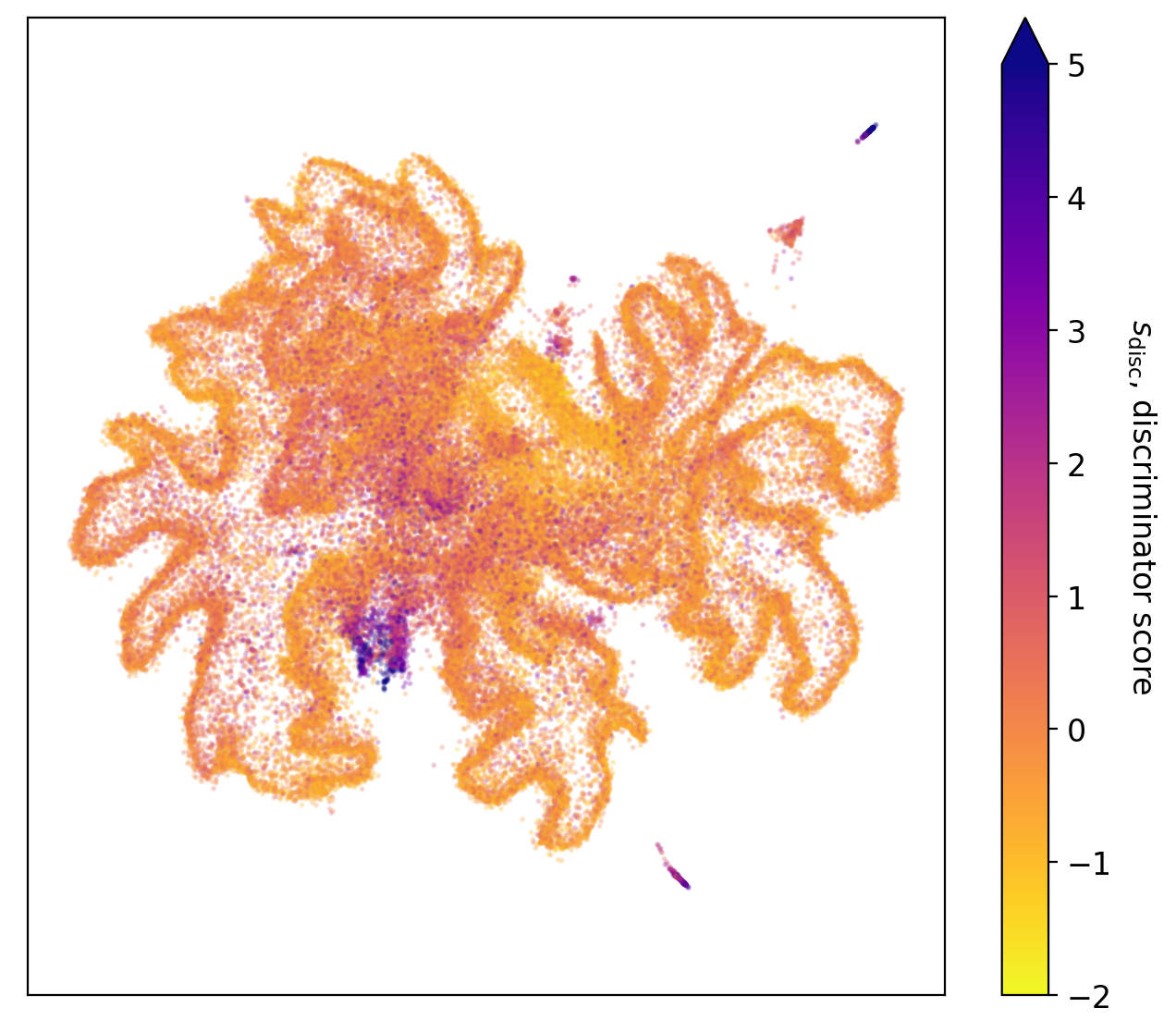}  
  \caption{Embedding with residual image pixels.}
  \label{fig:umap_100k_resids}
\end{subfigure}

\vspace{1em}

\begin{subfigure}{.48\textwidth}
  \centering
  \includegraphics[width=1\linewidth]{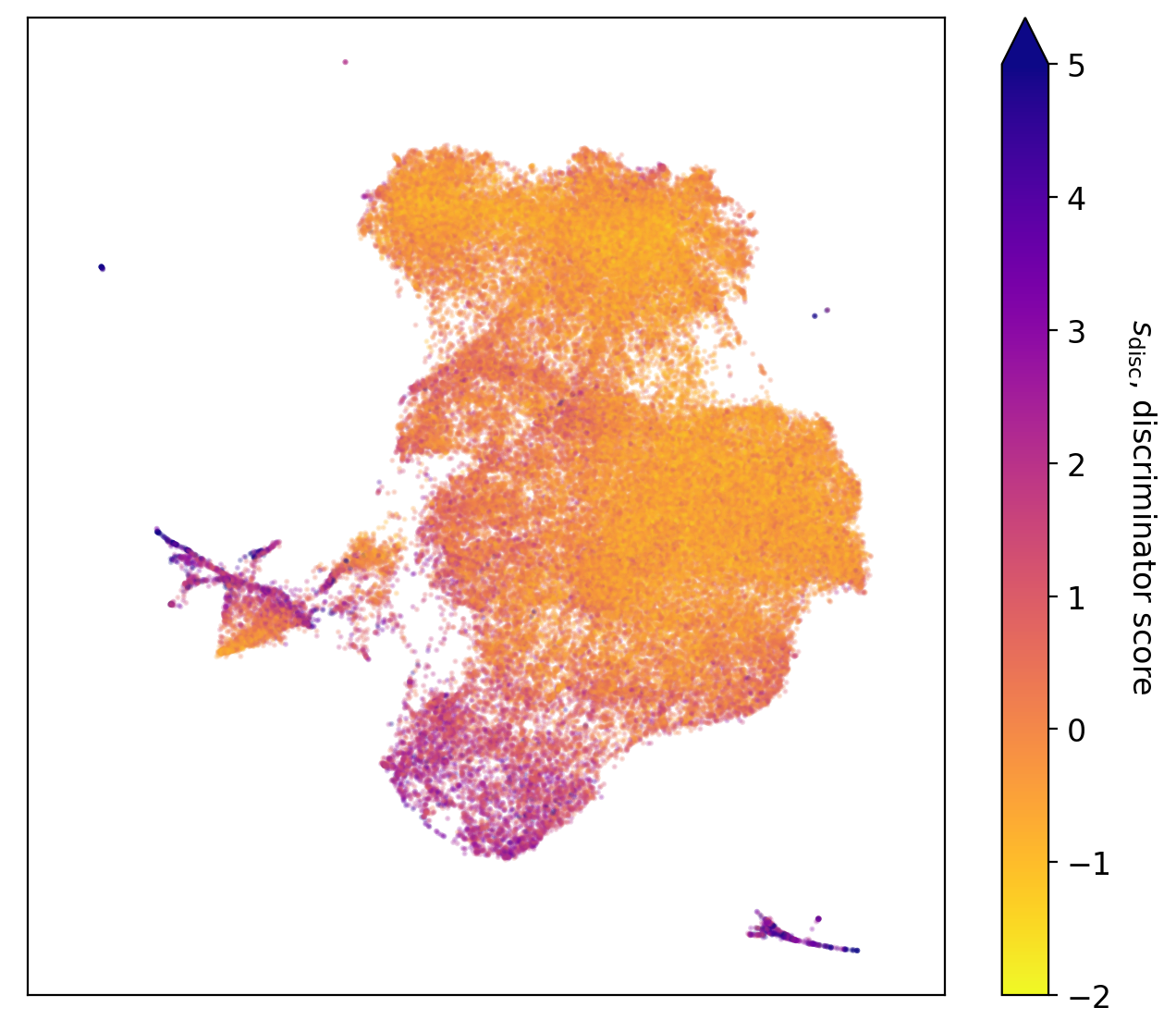}
  \caption{Embedding with autoencoded original images.}
  \label{fig:umap_100k_reals_auto}
\end{subfigure}
\begin{subfigure}{.48\textwidth}
  \centering
  \includegraphics[width=1\linewidth]{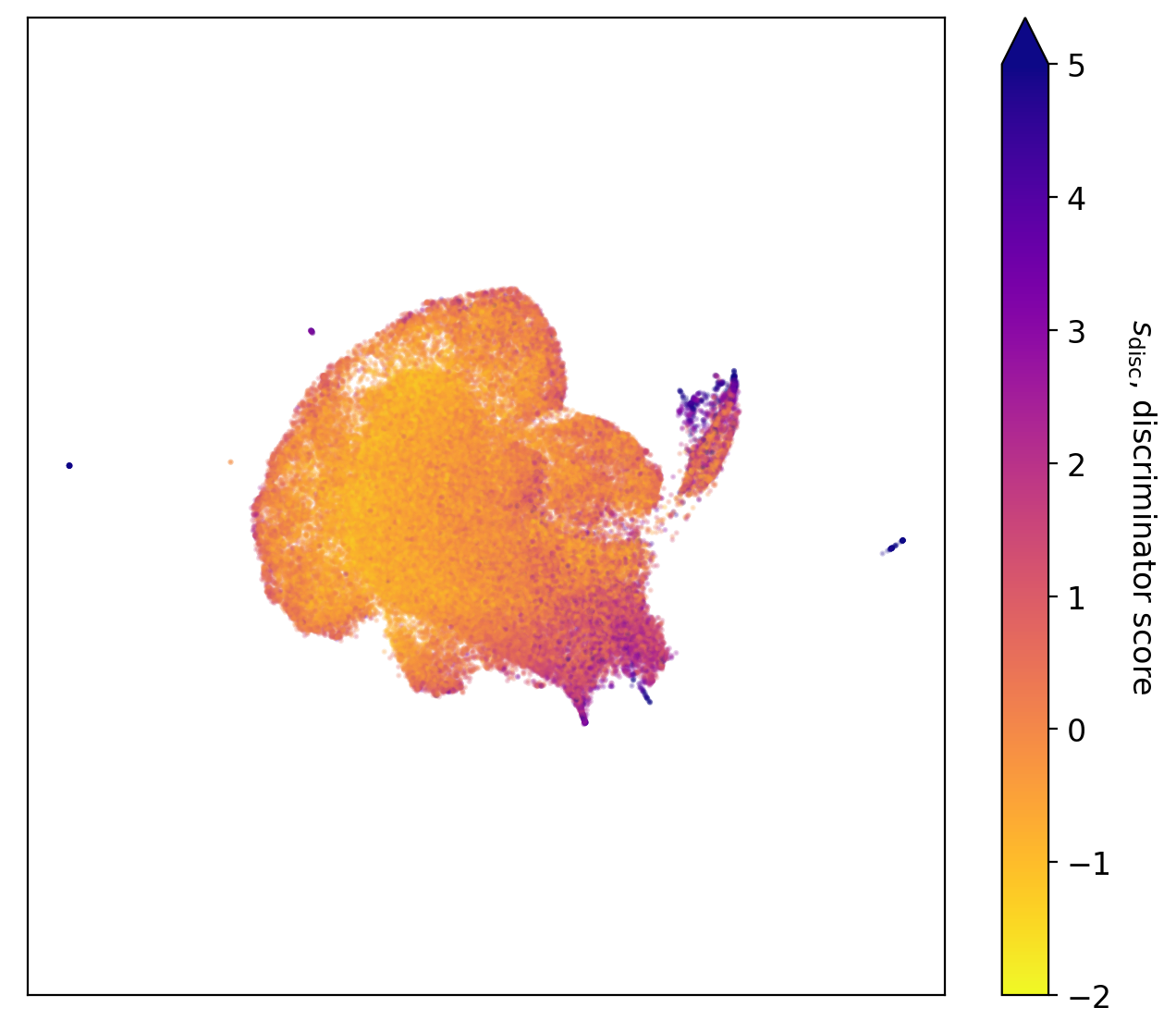}  
  \caption{Embedding with autoencoded residual images.}
  \label{fig:umap_100k_resids_auto}
\end{subfigure}

\caption{The anomalies in our sample visualized with a UMAP in two dimensions, with different features used for the UMAP embedding. The choice of autoencoded residual images (lower right) produces a UMAP distribution that is most strongly correlated with anomaly score.}
\label{fig:umap_100k}
\end{figure*}

We visualize the image distribution with a Uniform Manifold Approximation and Projection (UMAP, \citealt{McInnes2018}), a dimensionality reduction algorithm that maps the objects into a 2D representation.
This is useful for understanding the global properties of the distribution, and exploring the types of objects in the sample through their clustering UMAP-space.
We perform a UMAP embedding on a 100,000-object subsample of our data set, and look at the correlation between the UMAP and our WGAN-assigned anomaly scores.

We first perform an embedding directly on the 3-color image pixels; this is shown in Figure~\ref{fig:umap_100k_reals}.
There is a general trend with anomaly score, as well as significant structure in the distribution.
We next embed the residual image pixel values between the original images and the WGAN reconstruction, as we expect the residuals to contain information about the magnitude and type of anomaly; this is shown in Figure~\ref{fig:umap_100k_resids}.
We see an increased amount of structure, including a windy filamentary structure composed of lower-scoring objects, with the high-scoring objects somewhat clustered towards the center of the distribution.
The detailed structures of the low-scoring objects in these UMAPs suggest that the embeddings are dominated by less interesting pixel-level features.

In order to address this, we use a convolutional autoencoder (CAE) to reduce the dimensionality of these images, as described in Section~\ref{sec:cae}.
\new{We emphasize that while we use the same CAE for this application as we did for assigning CAE anomaly scores, as described in Section~\ref{sec:cae_bench}, here we use it for a totally different purpose: to aid in characterizing the anomalies found by the WGAN discriminator.}
The CAE finds a 64-dimensional latent-space representation of each original image, and separately of each residual image; these are much more compressed than the 27,648 dimensions of the images ($96 \times 96$ pixels in 3 color bands).

We first apply the UMAP to embed these autoencoded low-dimensional representations of the original images; this is shown in Figure~\ref{fig:umap_100k_reals_auto}.
We see that there is now a more coherent cluster containing most of the low-scoring anomalies, though there are high-scoring objects scattered throughout the distribution.
Finally, we show the result of embedding the autoencoded residual images in Figure~\ref{fig:umap_100k_resids_auto}. 
We obtain a coherent distribution with a very clear gradient in anomaly score.
This indicates that the CAE applied to the residual images is extracting information that is the most relevant to the WGAN-assigned anomaly score, as compared to the image pixels or the original images.
This provides support that our CAE technique, combined with our WGAN approach to anomaly detection, is useful for consolidating the information in the galaxy images relevant to their anomalousness.
In the next section, we show that this is useful for the further characterization of the high-anomaly sample to detect scientifically interesting anomalies.

\subsection{Characterization of Anomalies with the WGAN, CAE, and UMAP}

\begin{figure*}
\begin{subfigure}{0.68\textwidth}
  \centering
  \includegraphics[width=1\linewidth]{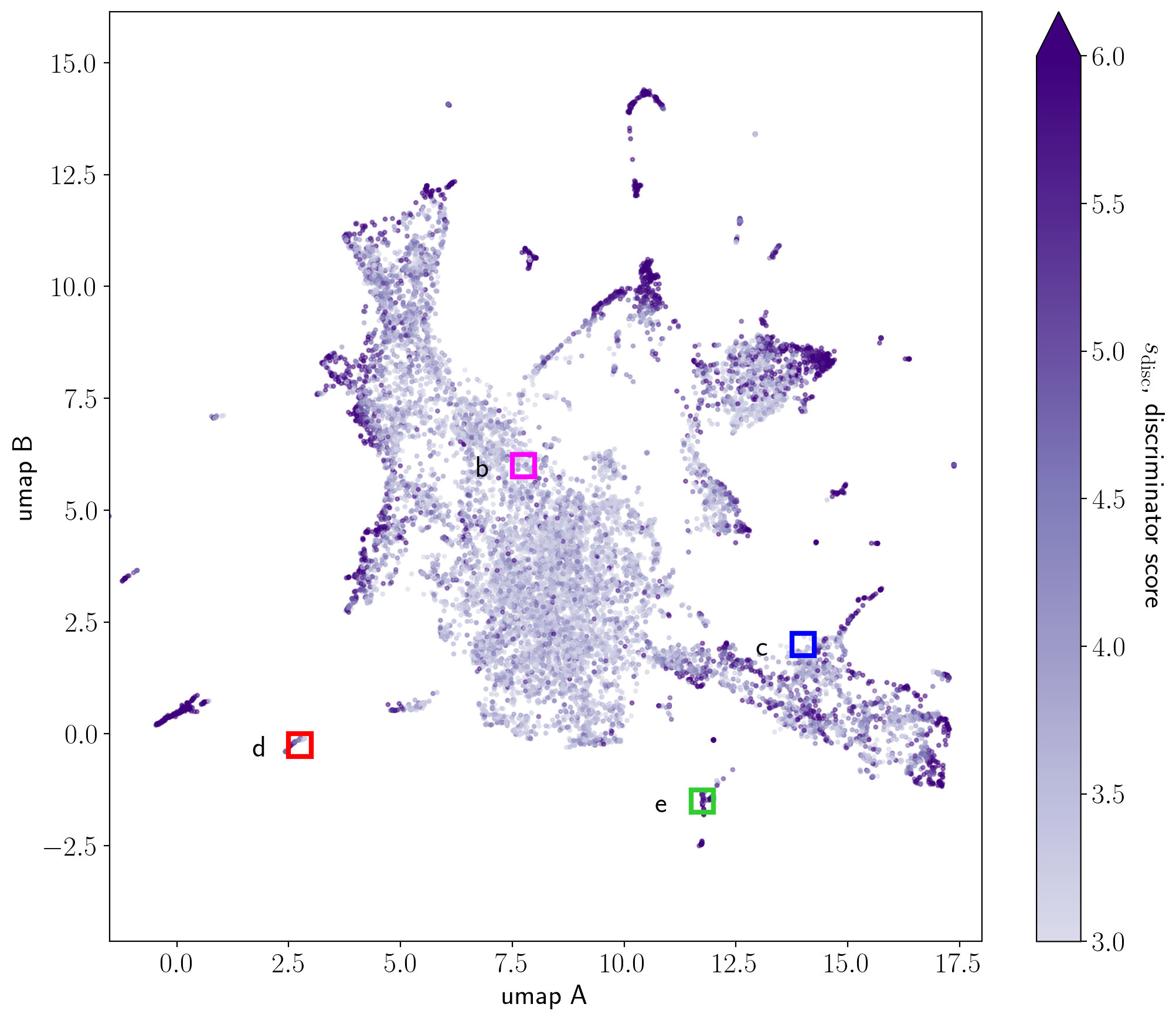}  
  \vspace{-2em}
  \caption{}
  \label{fig:umap_3sig_boxes}
\end{subfigure}

\begin{subfigure}{.48\textwidth}
  \centering
  \includegraphics[width=1\linewidth]{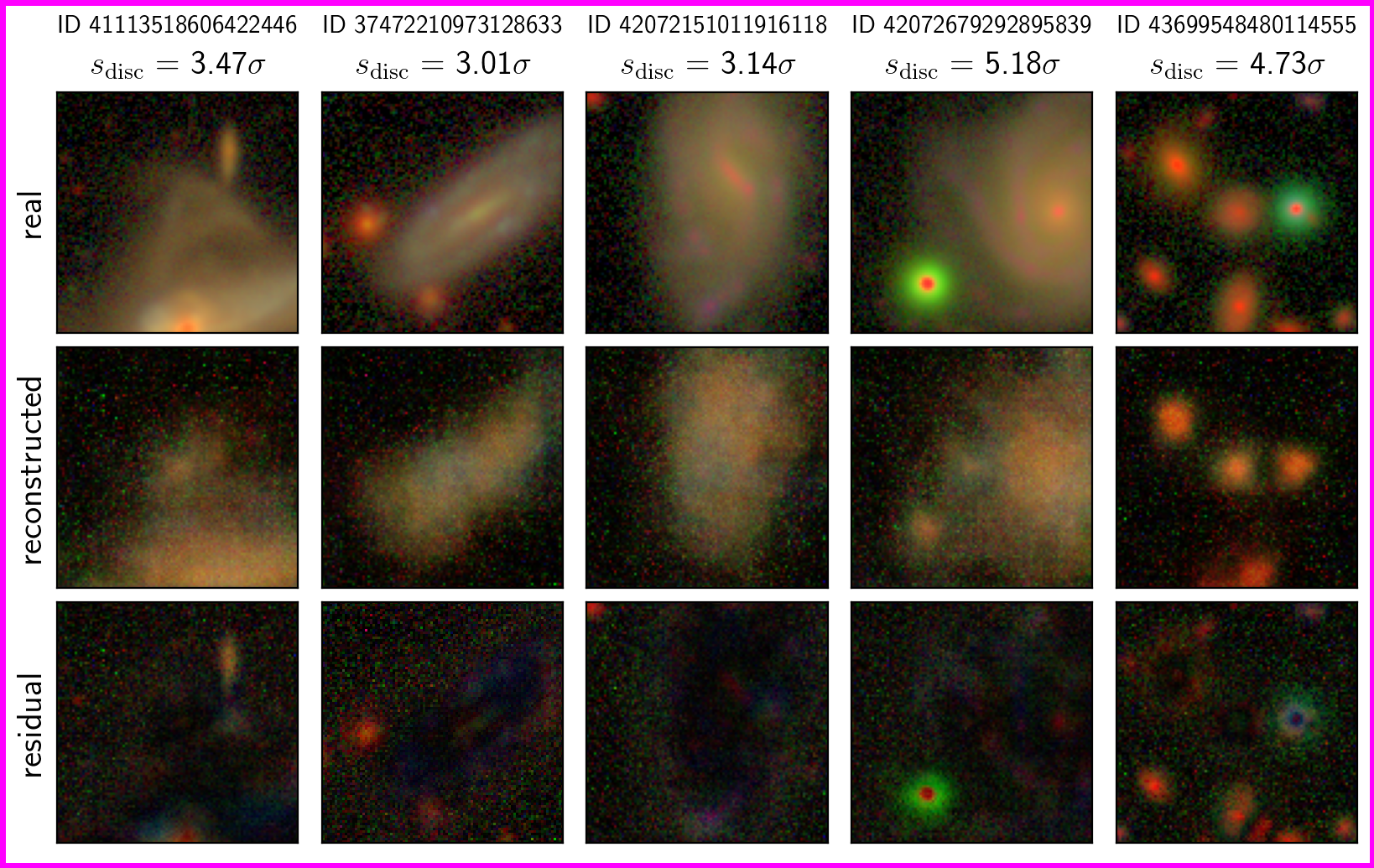}  
  \caption{}
  \label{fig:recons_magenta}
\end{subfigure}
\hspace{2em}
\begin{subfigure}{.48\textwidth}
  \centering
  \includegraphics[width=1\linewidth]{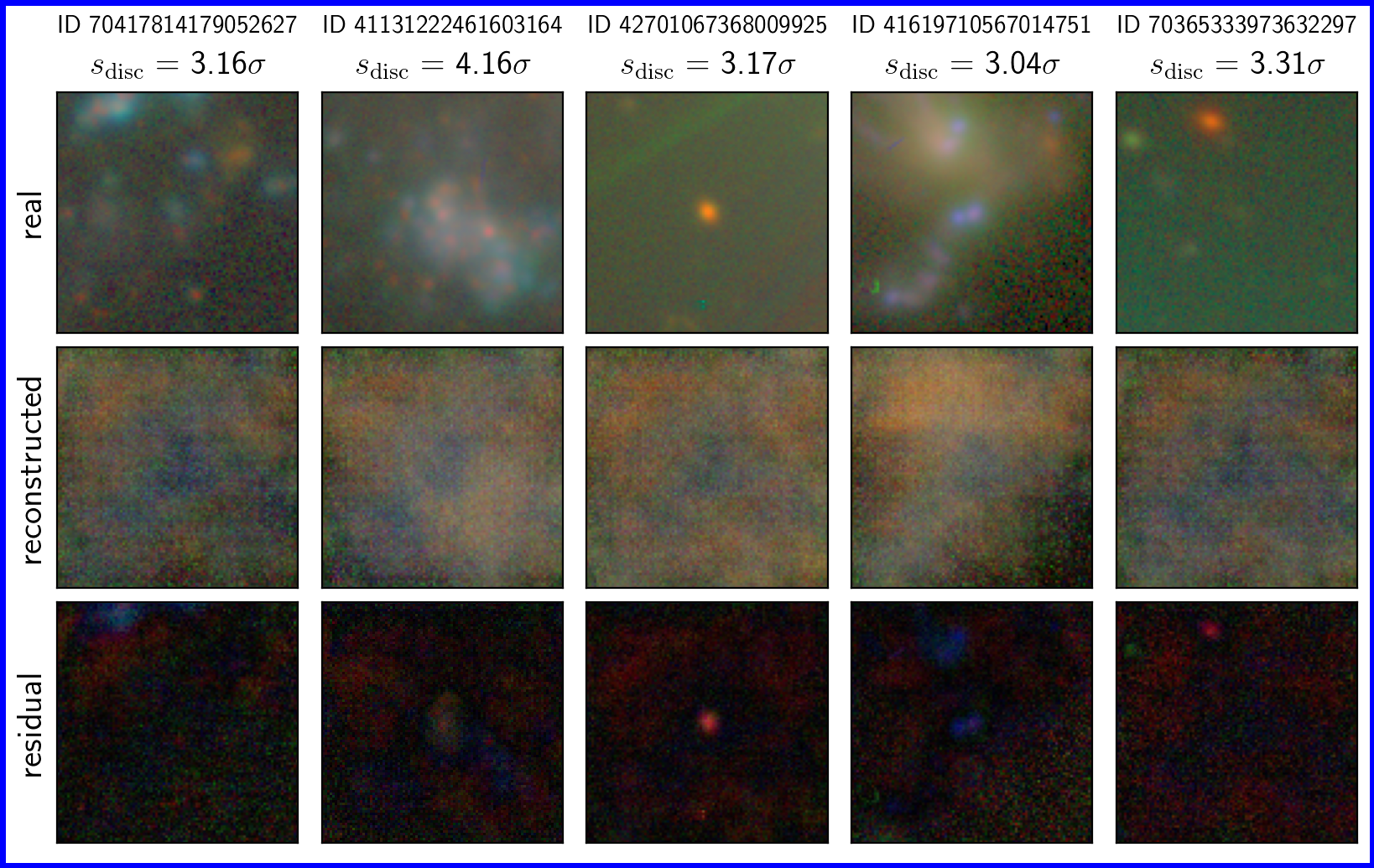}  
  \caption{}
  \label{fig:recons_blue}
\end{subfigure}

\begin{subfigure}{.48\textwidth}
  \centering
  \includegraphics[width=1\linewidth]{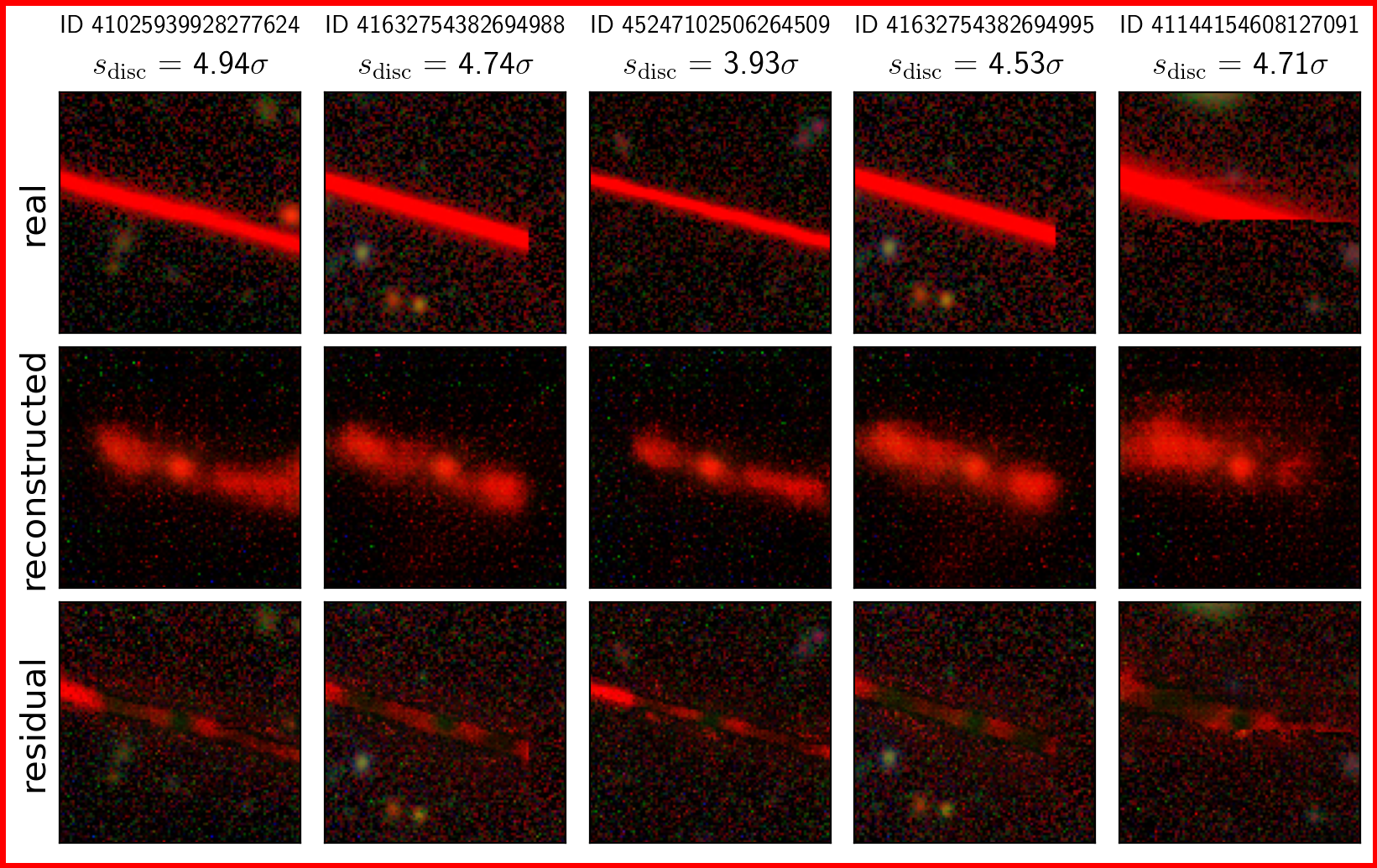}  
  \caption{}
  \label{fig:recons_red}
\end{subfigure}
\hspace{2em}
\begin{subfigure}{.48\textwidth}
  \centering
  \includegraphics[width=1\linewidth]{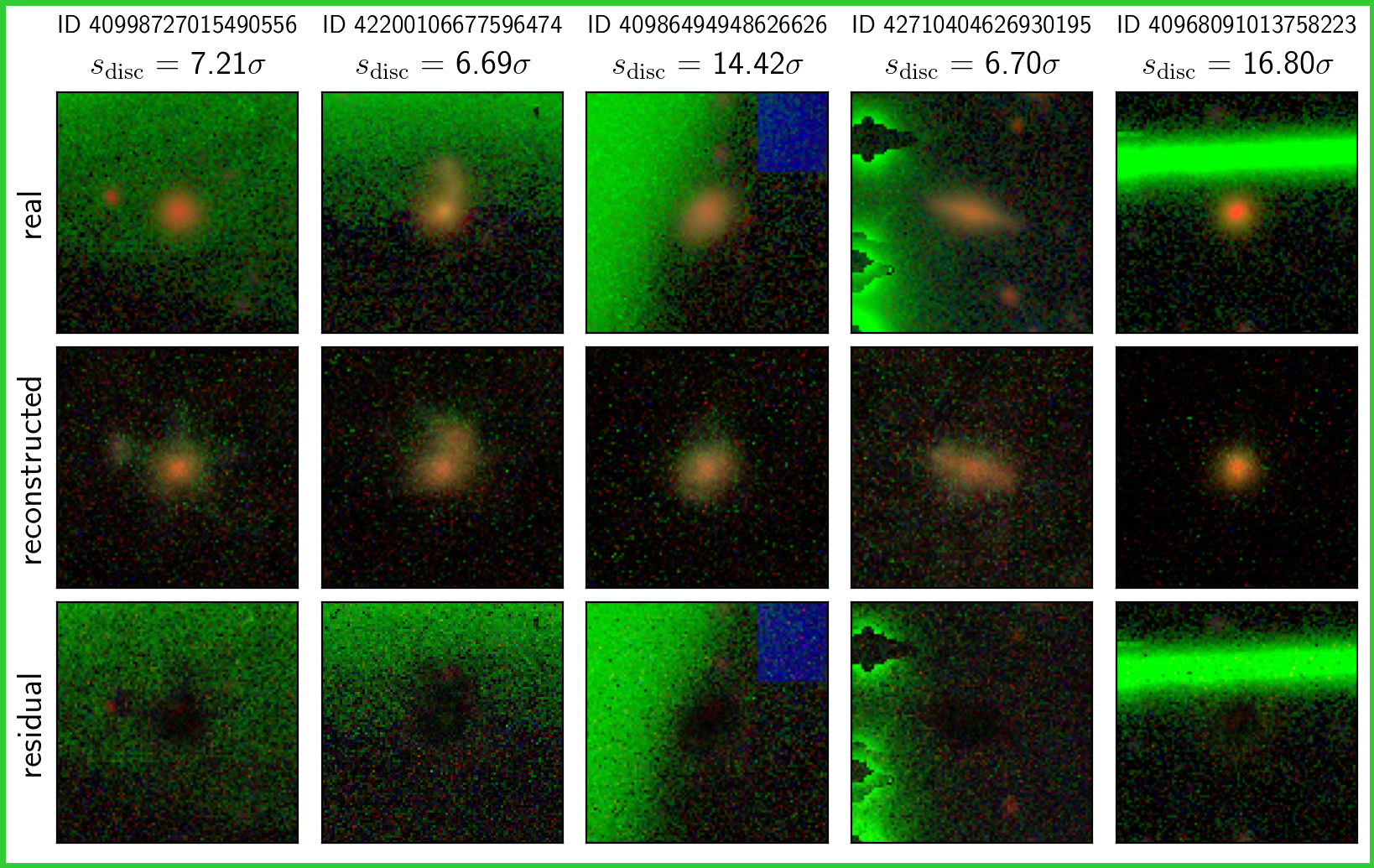}
  \caption{}
  \label{fig:recons_green}
\end{subfigure}

\vspace{0cm}
\caption{A visualization of our anomaly characterization method with our WGAN-based anomaly scores and autoencoded residual images. Panel (a) shows a UMAP embedding of all anomalies with  $\s{disc}>3\sigma$ above the mean, color-coded by anomaly score. Panels (b)-(e) show a random selection of galaxies in each of UMAP regions enclosed by the box of the corresponding color in panel (a). It is clear that different regions of the UMAP correspond to different types of anomalies, including both \new{interesting objects (as in (b) and (c)) and corrupted images (as in (d) and (e)).}}
\label{fig:boxes}
\end{figure*}

We use our WGAN-based anomaly score and our autoencoded residual images to explore and characterize the anomalous objects in our data set.
We perform a UMAP embedding of the autoencoded residuals for just the objects in our high-anomaly sample, as defined in Section~\ref{sec:sanom_dist}, as we are now interested in identifying the scientifically interesting anomalies among the high-scoring objects.
This is shown in Figure~\ref{fig:umap_3sig_boxes}.
The distribution shows a clear correlation with $\s{disc}$, with high-score structures around the edges of the distribution, while objects just passing the $3\sigma$ threshold are more evenly located in the center of the distribution.

In exploring the objects in this distribution, we find that the distribution reflects similarities in the objects and their residuals.
We demonstrate this by showing galaxies located in various regions of the UMAP, indicated by the colored boxes on Figure~\ref{fig:umap_3sig_boxes}.
Panels (b)-(e) of Figure~\ref{fig:boxes} show a random selection of galaxies from each of the boxes of the corresponding color.
The differences in the regions are clear visually.
Figure~\ref{fig:recons_magenta} shows images from the central region of the UMAP, which contain mostly extended galaxies with diffuse emission and little structure.
The residuals show large empty regions where the WGAN succeeded in reproducing the extended emission, with noise and background or foreground sources around the edges.
The nearness of these images in UMAP-space demonstrates that our approach successfully isolating the anomalous features and clustering based on these.
The images in Figure~\ref{fig:recons_blue} correspond to the edge of a high-scoring arm of the UMAP.
They all exhibit a strong blue-green color: some of these are due to noise, but others are bright blue sources that may be extreme star-forming regions.
We can see that the WGAN has difficulty reconstructing these complicated structures, and while it captures some of the blue color the reconstructions are dominated by diffuse yellow emission, which is more represented in the data.
We note that the attempted reconstructions are quite similar to each other, likely because there is only a small region of the WGAN's latent space that can produce images resembling these as they are far from typical training set images.
In addition to characterizing types of interesting anomalies, the UMAP is useful for separating out optical artifacts.
Figure~\ref{fig:recons_red} shows objects in a cluster far from the rest of the distribution, which contain red streaks aligned in the same direction, possibly due to satellites or other corruptions.
The WGAN struggles to reconstruct these; it attempts to line up multiple compact sources to approximate the line, as it was trained on majority compact sources, so the residuals all display similar features corresponding to gaps in the line.
The images in Figure~\ref{fig:recons_green}, from the cluster at the bottom of the UMAP, have bright green features due to corruption from nearby bright stars or passing satellites.
The WGAN reproduced the central object well but was unable to reconstruct the green features, as most of the data set does not contain these.

These sets of galaxies with distinct anomalous features demonstrate how our WGAN-based approach, combined with the CAE-enabled UMAP distribution, provides a useful way of characterizing anomalies.
We are able to disentangle anomalies that scored highly due to noise or saturation, such as those in panels (d) and (e), from those with a high score due to unusual galaxy morphology or color, such as in panels (b) and (c).
This indicates that our approach could be used to robustly filter out bad images at the pipeline level, in addition to identifying scientifically interesting anomalies in post-processing.
We built a custom visualization tool to interactively explore the UMAP space in more detail, based on a similar tool by \cite{Reis2021}; it can be accessed at \url{https://weirdgalaxi.es}.
We used this tool to perform a search for scientifically interesting anomalies; the results of this search are described in Section~\ref{sec:interesting}.

\subsection{Identified Interesting Anomalies}
\label{sec:interesting}

\begin{figure*}
\vspace{0cm}
\begin{subfigure}{.35\textwidth}
  \centering
  \includegraphics[width=1\linewidth]{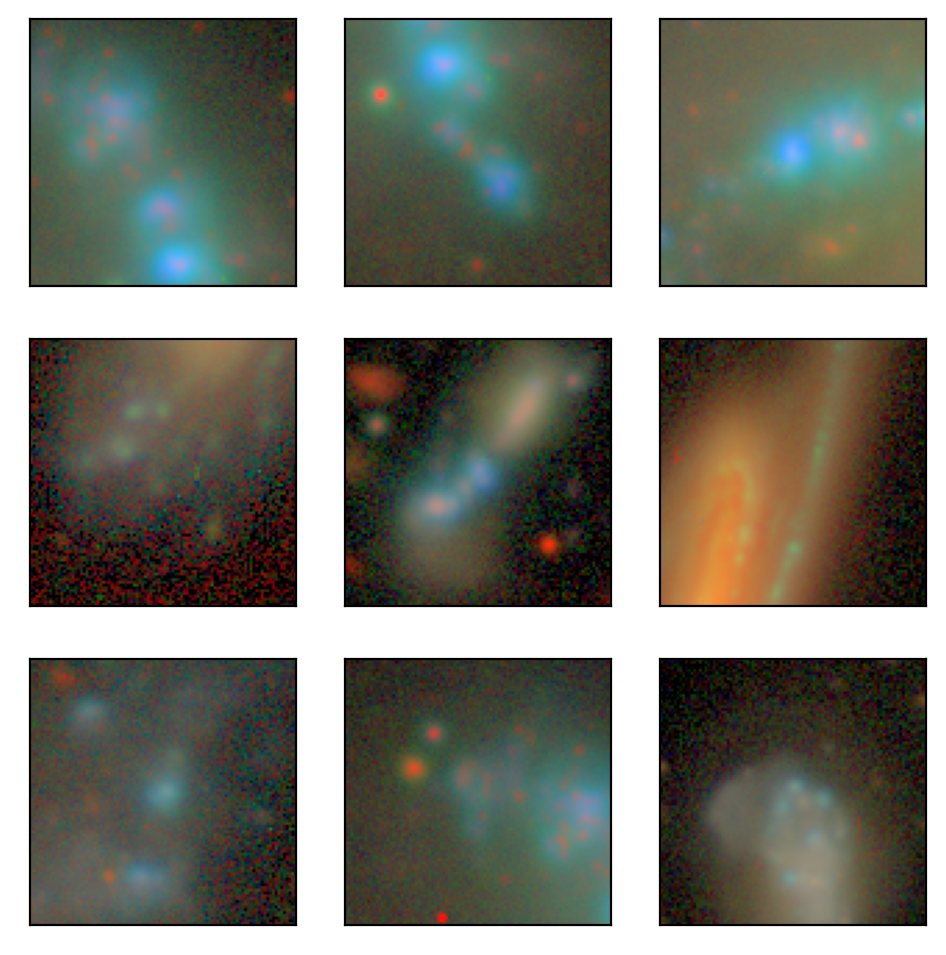}  
  \caption{Galaxies with extreme, blue star formation.}
  \label{fig:anom_bluesf}
\end{subfigure}
\hspace{2em}
\begin{subfigure}{.35\textwidth}
  \centering
  \includegraphics[width=1\linewidth]{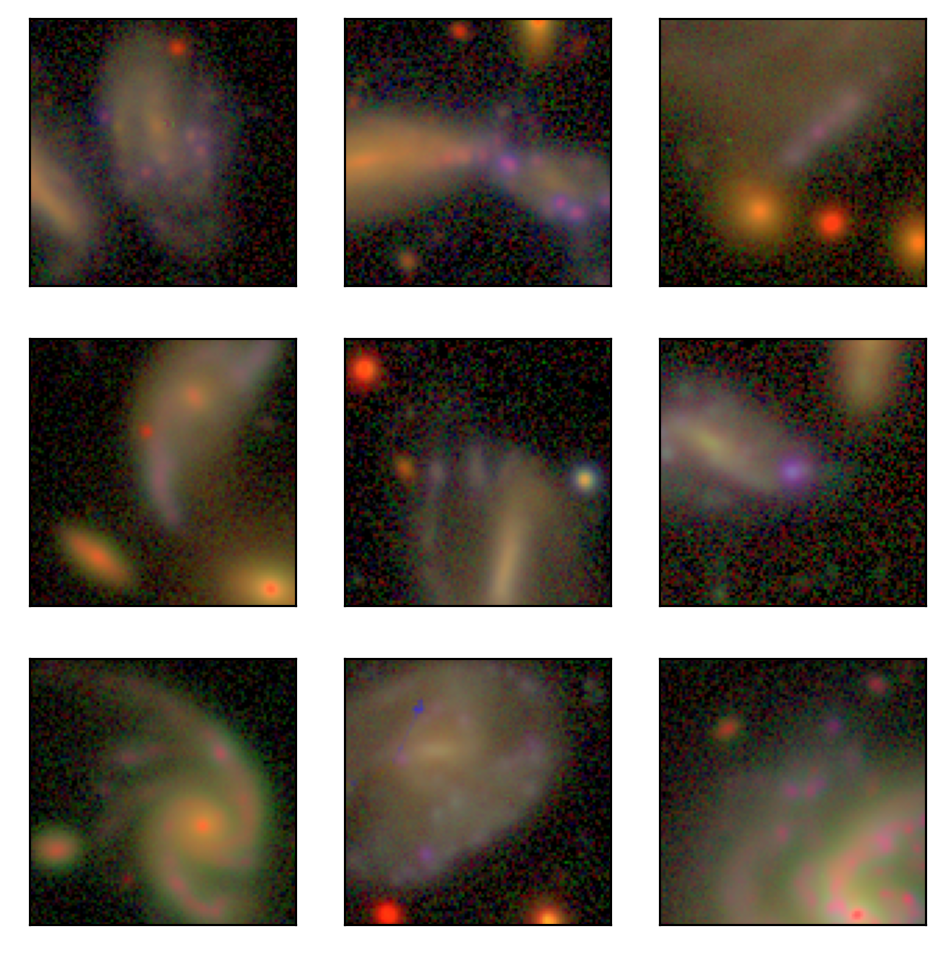}  
  \caption{Galaxies with purple active regions.}
  \label{fig:anom_purple}
\end{subfigure}
\vspace{1em}

\begin{subfigure}{.35\textwidth}
  \centering
  \includegraphics[width=1\linewidth]{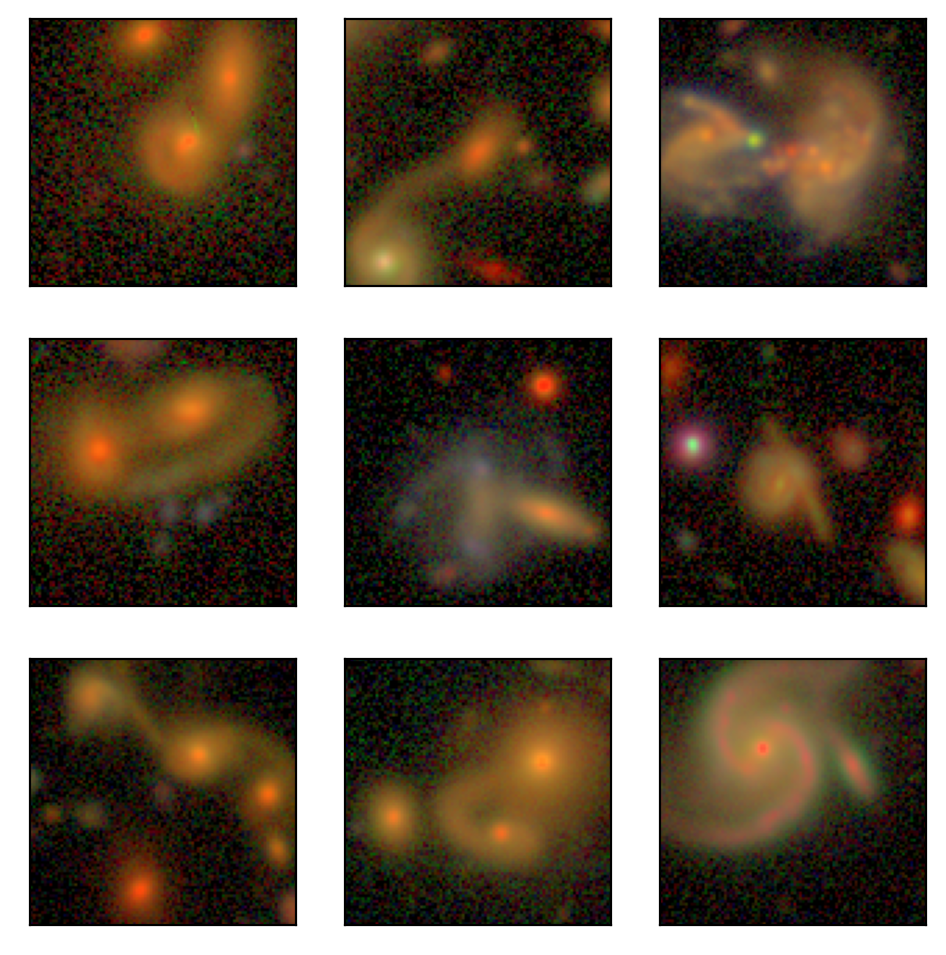}
  \caption{Galaxy mergers or potential mergers.}
  \label{fig:anom_mergers}
\end{subfigure}
\hspace{2em}
\begin{subfigure}{.35\textwidth}
  \centering
  \includegraphics[width=1\linewidth]{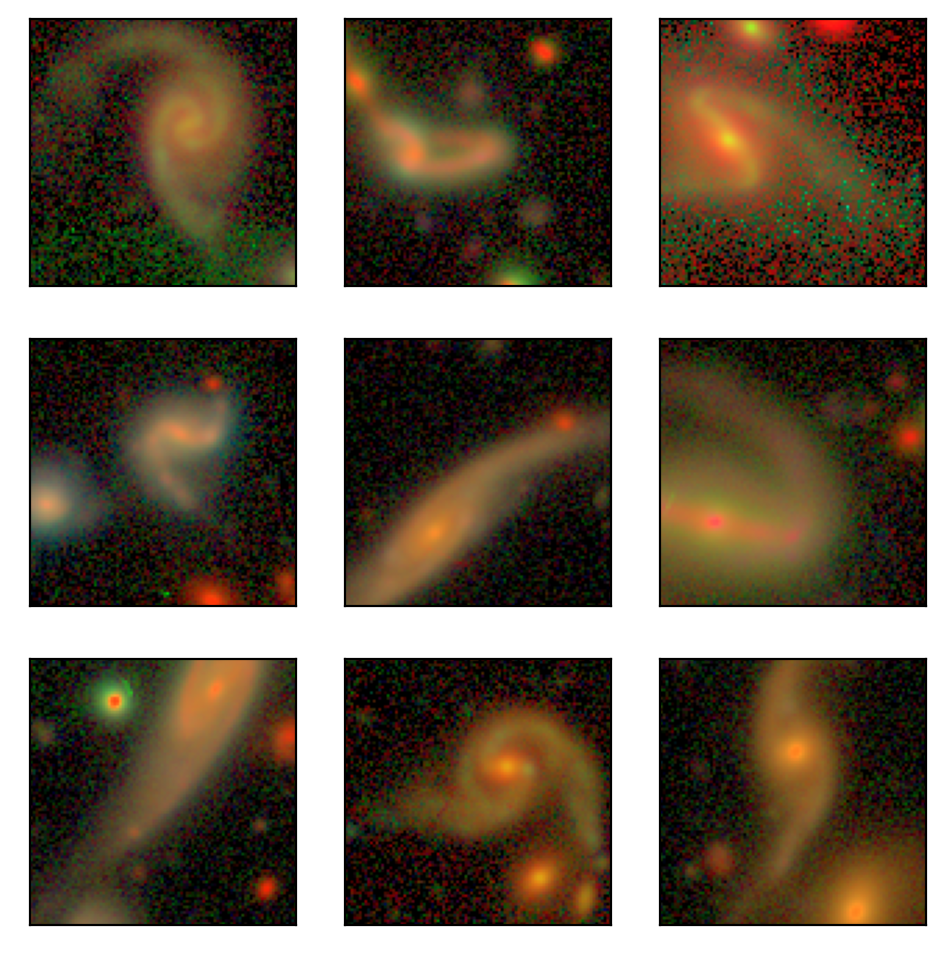}  
  \caption{Galaxies with tidal features.}
  \label{fig:anom_tidal}
\end{subfigure}

\vspace{1em}

\begin{subfigure}{.35\textwidth}
  \centering
  \includegraphics[width=1\linewidth]{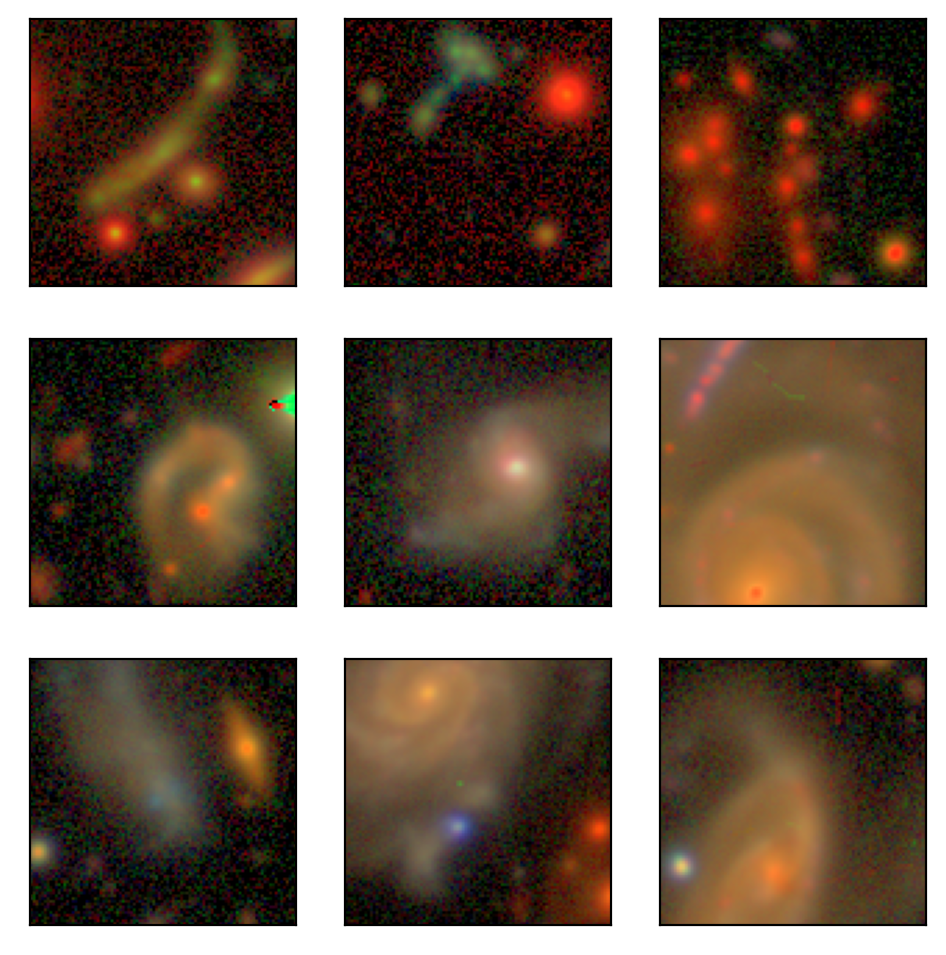}
  \caption{Images containing other potentially interesting anomalies.}
  \label{fig:anom_other}
\end{subfigure}
\vspace{0cm}
\caption{A selection of the interesting anomalies with scientific potential detected using our method.}
\label{fig:anomalies}
\end{figure*}

Using our approach combining WGAN-based anomaly scores and CAE-enabled characterization, we find a number of potentially scientifically interesting galaxy images.
A categorized selection of these is shown in Figure~\ref{fig:anomalies}.
We note that these categories were assigned by hand, and do not correlate very clearly with regions in UMAP-space; the identification and categorization required a decent amount of visual inspection using our interactive visualization tool.
That said, some of these objects do cluster in the UMAP, and in fact we found some of these objects by looking nearby previously identified images in the category of interest.

Figure~\ref{fig:anom_bluesf} shows galaxies with regions of intense blue emission, some more diffuse and some in discrete clumps; these indicate extreme star formation.
In fact, some of these objects have already been identified and followed up in the literature; for example, the leftmost and center image in the top row are part of MCG+00-25-010, a blue compact dwarf system that was found to have a metal-poor stellar population that powers extreme nebular emission \citep{Senchyna2017}.
We also find several extended galaxies with discrete intensely purple regions, shown in Figure~\ref{fig:anom_purple}.
This indicates strong emission in the $i$-band (red), which is from the H$\alpha$ line, as well as the $g$-band (blue), which could be H$\beta$ or [OIII] 4959 and 5007 lines; thus in certain redshift ranges, the purple regions would suggest star formation activity.
Figure~\ref{fig:anom_mergers} shows potential galaxy mergers, including one possible triple merger. 
We also detect many galaxies with tidal features due to gravitational interactions; a sample of these are shown in Figure~\ref{fig:anom_tidal}.
Finally, Figure~\ref{fig:anom_other} shows other anomalous images with potential scientific interest, including arcs of extended emission, a strangely shaped cluster of discrete sources, and bright compact blue and white objects in extended sources.
This last sample demonstrates the potential of our approach to identify ``unknown unknowns,'' which a more targeted anomaly detection method would likely have missed.

\new{All of the objects shown here were found in our $\s{disc} > 3\sigma$ sample.
Of these 45 images, only 11 of them were in the $\s{gen} > 3\sigma$ sample, and only 2 of them were in the $\s{CAE} > 3\sigma$ sample.
This confirms that the WGAN discriminator is well suited to finding interesting anomalies that other methods, especially the simpler CAE approach, cannot.
While it is likely that those samples would have contained some number of interesting anomalies that this discriminator sample does not, we showed in Figure~\ref{fig:score_effect} that more would be lost than gained in those cases.}

We performed follow-up spectroscopic observations of several of these objects to determine if they are indeed scientifically interesting.
\new{Our findings on one of these objects is presented in Appendix~\ref{sec:bluedot}.}

\section{Summary \& Conclusions}
\label{sec:conclusions}

In this work, we presented an approach combining a WGAN, CAE, and UMAP to detect anomalous images, and applied it to a sample of $\sim$940,000 objects in the Hyper Suprime-Cam survey.
We train a WGAN on the full data set in order to model the overall distribution of the data.
Data that are not well represented in the WGAN's latent space are identified as more anomalous with respect to the data set as a whole.
We quantify the degree of anomaly by setting the WGAN to find the closest representation of each object in its latent space and reconstruct the image.
We then assign scores based on the residuals between the original and the reconstruction, both the pixel-wise residual with the generator and a feature residual with the penultimate layer of the WGAN discriminator.

We found that the discriminator is more adept at identifying interesting anomalous images, while the generator tends to identify images that are anomalous due to noise or other optical artifacts.
\new{We compare the WGAN-based scores to a simpler anomaly detection method using a convolutional autoencoder (CAE).
We find that, while the CAE reconstructions have smaller residuals from the original images than the WGAN reconstructions, they are less indicative of interesting anomalies compared to the WGAN discriminator score.
This is likely due to the CAE producing smooth, noiseless reconstructions, and the WGAN discriminator's ability to model the true data distribution.}
We thus select a high-anomaly sample based on the WGAN discriminator scores to investigate in more detail, with a $\s{disc}>3\sigma$ cut resulting in 13,477 objects. 

One of the main difficulties with anomaly detection is determining which anomalies are scientifically interesting.
To address this, we augment our WGAN-based anomaly detection approach with a novel characterization method based on a CAE and a UMAP.
We use the CAE to reduce the dimensionality of the residual images, and use these lower-dimensional representations with a UMAP embedding to further cluster objects with high anomaly scores.

Using our approach, we identify numerous interesting anomalies with scientific potential, including galaxy mergers and galaxies with extreme star-forming regions.
\new{Of the 45 interesting anomalies we show from the WGAN discriminator score sample, only 2 would have been found with the CAE anomaly scores.}
We perform follow-up observations on some of these objects, and detail our findings on one of these.
The object is a compact blue source in a region of extended emission, which we find to likely be a metal-poor star-forming dwarf galaxy with unusual asymmetric emission lines, which we conclude is due to a spatially offset, extremely blue HII region.
This confirmed the scientific interest of an object detected with our approach, and demonstrates the potential for a synergy between machine-assisted anomaly detection methods and detailed observational follow-up.

We have publicly released a catalog of our full data set with our WGAN-assigned anomaly scores, together with our custom visualization tool for further exploring this data.
Our approach is flexible and can be applied to other data sets and data types.
The combination of the WGAN and CAE for unsupervised anomaly detection is scalable, reproducible, and removes spontaneity from the discovery process, making it ideal for extracting novel science from the increasingly large surveys of the coming decade.

\section*{Acknowledgements}

We gratefully acknowledge the Kavli Summer Program in Astrophysics for seeding this project; the initial work was completed at the 2019 program at the University of California, Santa Cruz.
This work was funded by the Kavli Foundation, the National Science Foundation, and UC Santa Cruz.
\new{N.R.’s work at Argonne National Laboratory was supported under the U.S. Department of Energy contract DE-AC02-06CH11357.}
K.S.F., A.L. and Y.L. are grateful for valuable insights on the interpretation of galaxy observations from Jenny Greene, Erin Kado-Fong, Kevin Bundy, Masami Ouchi, Kimihiko Nakajima, Yuki Isobe, Yi Xu, and Aaron Romanowsky.
K.S.F. also thanks Dezso Ribli, Lorenzo Zanisi, Itamar Reis, and the Flatiron Astrodata Group at the Center for Computational Astrophysics for helpful discussions. 
This material is based upon work supported by the National Science Foundation under Grant No. 1714610.
\new{We also thank the anonymous reviewer for very helpful comments.}

\section*{Data Availability}

The data underlying this article are available via the Hyper Suprime-Cam CAS Data Search tool at \url{https://hsc-release.mtk.nao.ac.jp/datasearch}.
The SQL query used to obtain the data set in this work is available at \url{https://github.com/kstoreyf/anomalies-GAN-HSC/blob/master/prepdata/hsc_pdr2_query.sql}.
A subset of the derived data, including anomaly scores for all the images, is available at \url{https://github.com/kstoreyf/anomalies-GAN-HSC}, along with the code used to process the data; the full data sets are available upon request.

\bibliographystyle{mnras}
\bibliography{Anomalies-HSC}

\appendix

\section{Follow-up Analysis of A WGAN-Detected Anomalous Galaxy}
\label{sec:bluedot}

\subsection{Source properties}

\begin{figure}
    \centering
    \includegraphics[width=0.4\textwidth]{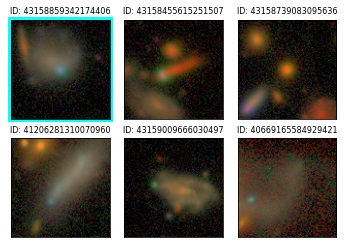}
    \caption{A sample of anomalous galaxies we detected with bright blue and purple sources within a region of more diffuse emission. We performed follow-up observations on the first three, and present a detailed analysis of the first object, HSC J095934+013707 (cyan border).}
    \label{fig:bluecores}
\end{figure}

We performed follow-up observations on several seemingly interesting sources that were found by our anomaly detection approach to determine whether they are indeed scientifically interesting.
Leading up to this, we first constructed a sample of galaxies with high anomaly scores from our main HSC sample that fell in the COSMOS field, where we had observing time on the DEIMOS spectrograph on the Keck II Telescope.
We used the characterization approach described here to examine the types of anomalies present.
One category of objects we found were images that contained bright blue compact sources situated towards the edges of larger regions of diffuse emission; these are shown in Figure~\ref{fig:bluecores}.
We performed follow-up observations of three of these, and obtained a good spectrum for one, with HSC ID 43158859342174406.
Here we present our analysis of this object as a demonstration of the scientific potential of our anomaly detection method.

The object has coordinates RA=09:59:34.060, dec=+01:37:07.84; we refer to it as HSC J095934+013707 (it is also in the COSMOS catalog, labelled as COSMOS 244571).
The association between the blue compact source and the diffuse emission is unclear from visual inspection; they may be two associated galaxies, or a single galaxy with a bright feature, or unrelated overlapping objects.
The object is in the COSMOS catalog and is computed to have a mass of $10^{7.9} M_\odot$, based on 30-band SED fitting, and a redshift $z=0.0320$. 

\subsection{Spectroscopic analysis of source}

\begin{figure*}
    \centering
    \includegraphics[width=0.9\textwidth]{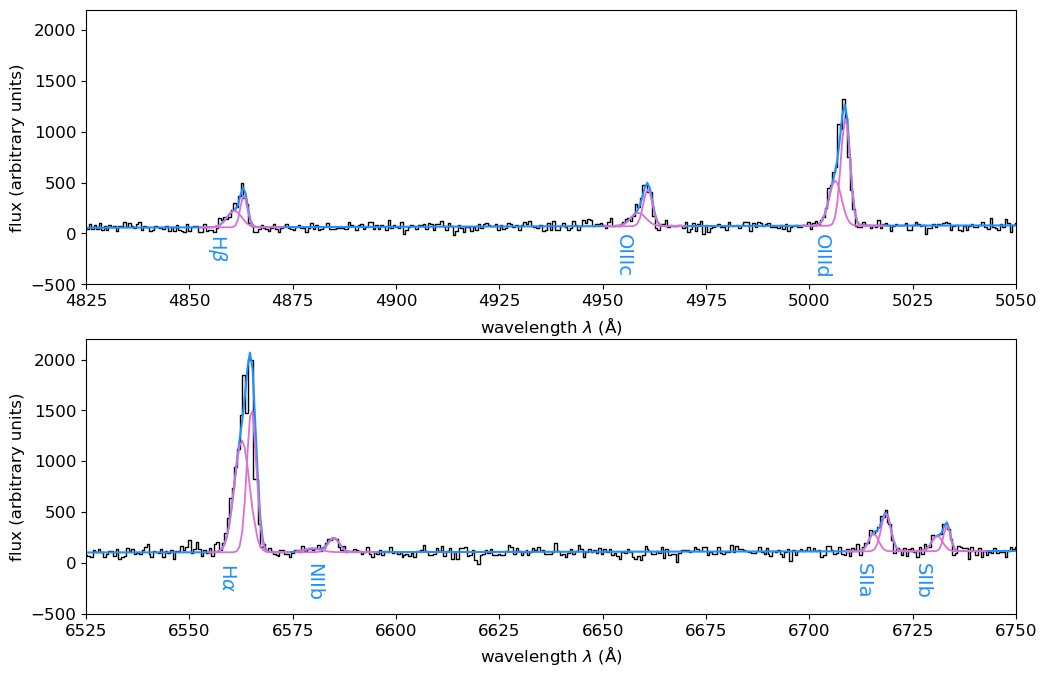}
    \caption{Spectrum of our interesting source, HSC J095934+013707 (black), converted to rest wavelength. The H$\alpha$ region is shown in the top panel the and H$\beta$ region in the bottom panel. We plot the best double Gaussian fits, with the individual components in pink and the total in blue.}
    \label{fig:spectra}
\end{figure*}

We placed the slit centered on the compact source and crossing the diffuse emission. 
The spectrum we obtained is shown in Figure~\ref{fig:spectra}, zoomed in on regions with visible emission lines.
We compute a redshift of $z=0.03221$ from the H$\alpha$, H$\beta$, and OIII lines, very close to that from the COSMOS catalog.
There is only one detectable redshift, so we conclude that the compact and diffuse components are at the same cosmological redshift.
We do see a slightly offset component of faint emission that may indicate a spatially separated region in the system.
The diffuse emission has a diameter of $\sim4.1"$, so given this redshift, it has a spatial extent of $\sim2.7$ kpc.
The spectrum, which is the combined signals of the blue and diffuse sources, shows strong H$\alpha$, very low NII, moderate H$\beta$, and moderate OIII lines, as well as moderate SII lines. 
This indicates that the source is highly star-forming and metal-poor.

All of the emission lines exhibit a clear asymmetry, with a blueshifted tail.
We perform double Gaussian fits to each of the emission lines; these are much better fits than those with single Gaussians.
For most of the lines, but particularly H$\alpha$ and OIIId, there is a large component on the blue side.
The strength of this asymmetry is highly unusual, though some asymmetry has been observed in dwarf galaxies with gas outflows or inflows in MaNGA \citep{Wylezalek2020, Avery2021}, as well as in some extremely metal-poor dwarf galaxies (EMPGs) in the EMPRESS sample \citep{Kojima2019}.
This initially suggests that our source is a dwarf galaxy with a gaseous outflow; we investigate this possibility by looking at the components of the flux fit.

The two components of the double Gaussian fits are separated by a relative velocity of $\sim180$ km/s; this could imply either two separate but interacting galaxies, or one galaxy with an inflowing or outflowing source.
The higher-wavelength component has a larger flux for all lines; this suggests that it corresponds to the bright blue source (also
consistent with the spatial distribution of
the emission in the 2D spectrum), while the lower-wavelength component corresponds to the fainter, more diffuse galaxy.
This is corroborated by a previous spectrum of the object taken as part of the DEIMOS 10K survey: this spectrum shows the same emission lines but they are completely symmetric, and centered on the lower-wavelength components. 
The slit for this observation was, most likely, at an angle that did not cross the blue source, and only captured the diffuse region.
This component analysis disfavors the interpretation of a gaseous outflow, as the blue source is redshifted with respect to the associated galaxy.
Rather, this understanding, combined with the strong emission lines, suggests that the blue source is a strong HII region, associated with a more diffuse but still star-forming dwarf galaxy system.

The spatial and dynamical relationship between the two components remains somewhat unclear.
The blue source may be in front of the diffuse region, moving towards it, or behind it and moving away from it; in the latter case, we would still expect to see the strong blue source shining through the diffuse galaxy image if that galaxy lacks significant dust.
Indeed, we find that the Balmer decrement is $H\alpha/H\beta = 2.80$, very close to the expected value of 2.86 for star-forming galaxies without dust obscuration (though this value does not vary significantly with galaxy properties, \citealt{Osterbrock2006}).
The lack of dust does fit with our understanding that this region is a dwarf galaxy, which tend to have little dust.
We might expect to find HII regions like the blue source off-center in dwarf galaxies; an event in the object's history could have triggered extreme star formation that ionized the gas.
Further, the asymmetry of the diffuse region suggests possible disruption from interaction in the past.
That said, the large velocity offset between the blue source and the diffuse galaxy mean that we cannot be sure about their interaction history or current dynamics.

We compute the fluxes of the emission lines for the individual and combined components from the Gaussian fits.
We compute the oxygen abundance of the combined object with three different methods, and find values in the range $12+\mathrm{log(O/H)} = 7.92-8.16$.
We separately compute the metallicities of the lower-wavelength and higher-wavelength components. 
The lower-wavelength component has $12+\mathrm{log(O/H)} = 7.71-8.13$, while the higher-wavelength component has $12+\mathrm{log(O/H)} = 8.03-8.24$.
For all of these cases, the source is low metallicity; however, it is not as low as EMPGs, which are defined as having $12+\mathrm{log(O/H)} < 7.69$, less than 10\% of the solar value.
The metallicity is also higher than the typical range for some other types of galaxies that bear some resemblance to ours, including blueberry galaxies \citep{Yang2017}, green pea galaxies \citep{Cardamone2009}, and Ultra Blue Compact Dwarfs \citep{Corbin2006}.
These metallicities do fall in the range of Luminous Blue Compact Galaxies (e.g. \citealt{Hoyos2007}), which tend to be low-mass, aligning with our mass measurement.

We see that the higher-wavelength component, which we take to be the blue HII region, has higher metallicity than the lower-wavelength component (diffuse region) for all ways of estimating the metallicity.
This is quite unexpected, as HII regions are typically metal-poor compared to their surrounding galaxy.
The source could be an HII region that is self-enriched, making it higher metallicity compared to its host galaxy, which is possible though rare \citep{Kroger2006}. 

\begin{figure}
    \centering
    \includegraphics[width=0.48\textwidth]{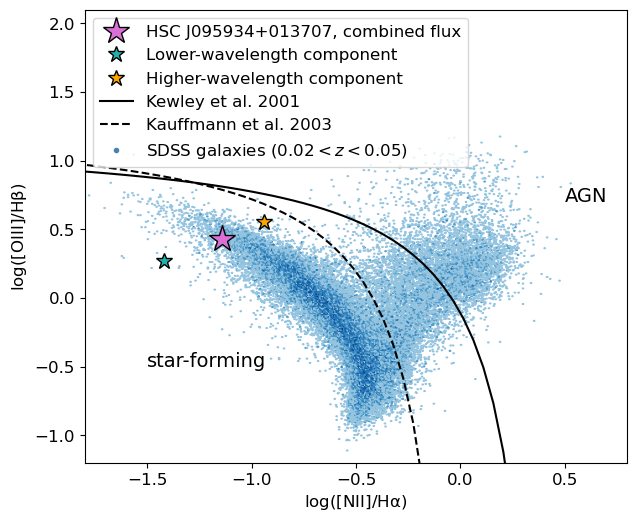}
    \caption{BPT diagram for our source, HSC J095934+013707. We show the position of the combined flux measurement (pink star), and the positions with the flux measured just from the lower-wavelength component (teal star) and the higher wavelength component (orange star). For comparison, we show the position of SDSS galaxies with similar redshift (blue dots).}
    \label{fig:bpt}
\end{figure}

We show the position of our source on the Baldwin, Phillips \& Terlevich (BPT, \citealt{Baldwin1981}) diagram in Figure~\ref{fig:bpt}.
We show both the individual components and the combined flux positions; we also plot the \cite{Kauffmann2003} classification between star-forming and composite galaxies, and the \cite{Kewley2001} starburst limit.
For comparison, we show a sample of galaxies from the Sloan Digital Sky Survey (SDSS, \citealt{York2000}) with $0.02<z<0.05$ (close to our galaxy's redshift of $z=0.03221$), only selecting objects with S/N > 3 for all the four emission lines in the BPT diagram, and $b/a > 0.5$ to filter out edge-on galaxies.
Both components of our source are solidly in the star-forming region, in both methods of classification; this rules out the possibility of an AGN.
The fact that lower-wavelength component has smaller line ratios compared to the higher-wavelength component aligns with metallicity comparison above.
However, we note that these have large error bars due to degeneracies in the double Gaussian fit, so the difference in metallicity may not be significant.
The combined-flux emission ratios place the object at the star-forming edge of the SDSS sample, though it is not an extreme outlier.
The lower-wavelength component does fall outside the bulk of the distribution of SDSS galaxies.

\begin{figure}
    \centering
    \includegraphics[width=0.43\textwidth]{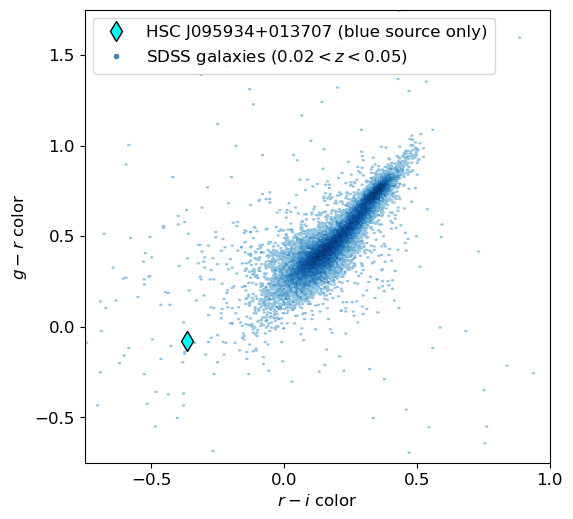}
    \caption{Color-color diagram for our source, HSC J095934+013707. The color is computed for just the bright blue source. A sample of SDSS galaxies with similar redshift are shown for comparison.}
    \label{fig:color-color}
\end{figure}

Finally, we show the color-color diagram for our source, compared to the same SDSS sample (Figure \ref{fig:color-color}).
The color is meaured from the CModel flux magnitudes of just the compact blue source.
We see that the source has an extremely blue color, more so than most of the SDSS galaxies.
This further supports our conclusion that the blue source is a very extreme HII region.

Based on this analysis, we conclude that the source is most likely a metal-poor star-forming dwarf galaxy with an associated self-enriched HII region, which is spatially distinct from the host galaxy. 
That said, there are still open questions about this object; we encourage follow-up observations of this and similar sources to better understand this type of system.
In any case, we did confirm that this is indeed a very interesting object: the asymmetry of the emission lines is a rare feature, the compact source is extremely blue, and the relative metallicity of the system components is not easily explained.

\label{lastpage}
\end{document}